\documentclass[12pt]{spieman}  
\usepackage{amsmath,amsfonts,amssymb}
\usepackage{graphicx}
\usepackage{setspace}
\usepackage{tocloft}
\usepackage{lineno}
\usepackage[numbers,sort&compress]{natbib}
\usepackage{multirow}
\usepackage{lscape}
\usepackage{hyperref}
\usepackage[symbol]{footmisc}

\title{Tracking turbulence for wind-driven halo removal using MagAO-X pyramid wavefront sensor images\footnote[7]{This paper includes data gathered with the 6.5 meter Magellan Telescopes located at Las Campanas Observatory, Chile.}}

\author[a,b,*]{Jay K. Kueny}
\author[a]{Jared R. Males}
\author[e]{Alycia J. Weinberger}
\author[a]{Laird M. Close}
\author[d]{Joseph D. Long}
\author[a,b]{Joshua Liberman}
\author[f,a]{Sebastiaan Haffert}
\author[a,b,c]{Eden McEwen}
\author[a]{Maggie Y. Kautz}
\author[g,a,b]{Olivier Guyon}
\author[h]{Logan Pearce}
\author[a,c]{Parker T. Johnson}
\author[a,b]{Katie Twitchell}
\author[a,c]{Jialin Li}
\author[i]{Alex Hedglen}
\author[j]{Avalon Gower}
\author[a]{Warren Foster}
\author[k]{Jhen Lumbres}
\author[l]{Lauren Schatz}

\affil[a]{Steward Observatory, University of Arizona, 933 N Cherry Ave, Tucson, AZ 85721, USA}
\affil[b]{James C. Wyant College of Optical Sciences, University of Arizona, 1630 E. University Blvd., Tucson, AZ 85721, USA}
\affil[c]{National Science Foundation Graduate Research Fellow}
\affil[d]{Center for Computational Astrophysics, Flatiron Institute, 162 5th Ave, New York, NY}
\affil[e]{Earth and Planets Laboratory, Carnegie Institution for Science, 5241 Broad Branch Road NW, Washington, DC 20015-1305}
\affil[f]{Leiden Observatory, Leiden University, PO Box 9513, 2300 RA Leiden, The Netherlands}
\affil[g]{Subaru Telescope, National Observatory of Japan, Hilo, HI}
\affil[h]{Department of Astronomy, University of Michigan, Ann Arbor, MI}
\affil[i]{Northrop Grumman Corporation, 600 South Hicks Rd, Rolling Meadows, IL}
\affil[j]{Draper Laboratory, 555 Technology Square, Cambridge, MA}
\affil[k]{Northrop Grumman, Pasadena, CA}
\affil[l]{Starfire Optical Range, Kirtland Air Force Base, Albuquerque, NM}

\newcommand{\hr}{HR~4796A}
\newcommand{\farcsec}{\hbox{$.\!\!^{\prime\prime}$}}

\newcommand{\zband}{$z'$-band}
\newcommand{\iband}{$i'$-band}
\newcommand{\rband}{$r'$-band}
\newcommand{\gband}{$g'$-band}
\newcommand{\adi}{KLIP-ADI}

\cftpagenumbersoff{figure}
\cftpagenumbersoff{table}
\begin{document}
\maketitle

\begin{abstract}
The wind-driven halo (WDH) is a persistent, low spatial frequency noise artifact that arises due to the servo-lag error inherent to all adaptive optics (AO) instruments. Spatial filtering may be employed to overcome this artifact, however, filtering out the WDH while simultaneously preserving signal from an extended astrophysical object of interest is exceptionally challenging. Additionally, since the WDH changes in intensity and position angle through an observation, data-driven algorithms (e.g., KLIP) that are commonly used to subtract the starlight need to be overly-aggressive to remove both the static and dynamic noise components. Since wavefront sensors (WFSs) continuously track the closed-loop residual wavefront error, WFS telemetry presents the ideal resource for combating this type of noise artifact through postprocessing. Using archival WFS telemetry from MagAO-X, which is the ``extreme" AO instrument for the 6.5-meter Magellan-Clay telescope, we demonstrate a novel workflow for WDH reconstruction and removal in individual coronagraphic science images. MagAO-X is equipped with a pyramid WFS capable of recording wavefront telemetry at a high-cadence which is saved during data acquisition. Given this, we detail how our WFS data processing pipeline, \texttt{windsoCC}, cross-correlates the recorded closed-loop wavefront to measure the wind vectors of several turbulent layers of the atmosphere above Las Campanas Observatory. We then make use of the wind parameters learned through \texttt{windsoCC} to reconstruct the WDH footprint by leveraging a parametric model. Notably, we demonstrate a dramatic improvement in object recovery using on-sky MagAO-X images of the disk around HR~4796A at visible wavelengths.
\end{abstract}


{\noindent \footnotesize\textbf{*}Jay K. Kueny, \linkable{jkueny@arizona.edu} }

\begin{spacing}{2}   

\section{Introduction}
\label{sec:intro}

Expanding our knowledge on the close-in, circumstellar environments of nearby stars is critical for understanding planet formation and dynamics associated with extrasolar systems at various evolutionary stages. Large, ground-based telescopes, when assisted with adaptive optics (AO), present an ideal resource for achieving this goal given the diffraction-limited imaging capabilities of apertures in the 6 to 10 meter class. In particular, studying the region within a few hundred mas of the host star is of interest for learning more about planet formation as that is where gaseous giant planets are most likely to be found (e.g., \cite{nielsenGeminiPlanetImager2019}; \cite{viganSPHEREInfraredSurvey2021}). There is also value in the study of circumstellar debris disks as they are comprised of the leftover building blocks of planets \cite{hughesDebrisDisksStructure2018}. Imaging the light scattered by the dust in a circumstellar disk, especially the scattering at the minor axis, is most valuable for inferring its physical properties (\cite{minMultiwavelengthOpticalProperties2016}, \cite{arnoldStumblingPlanetaryBuilding2022}, \cite{kuenyMultibandStudyHR2026}). However, the angular separation of the minor axis of a circumstellar disk decreases with increasing disk inclination making the systems with the most-coveted degree of forward scattering challenging to image. It is these aforementioned science goals that spotlight the high-contrast imaging community's efforts to minimize scattered starlight while maximizing sensitivity to close-in planets and dust.

One persistent source of noise in the AO correction zone is the wind-driven halo (WDH; \cite{cantalloubeOriginAsymmetryWind2018}; \cite{madurowiczCharacterizationLemniscateAtmospheric2018}; \cite{madurowiczAsymmetriesAdaptiveOptics2019}) which manifests as a butterfly-shaped artifact spanning across this region of the image. It arises due to the servo-lag error inherent to all AO instruments. In short, because of the latency between wavefront sensor (WFS) measurement and application of the correction on the deformable mirror (DM), fast-moving turbulence causes the correction to become stale before it is applied; the jet stream is often a significant contributor to the WDH artifact seen in the science focal plane. It is this constant inability of the AO system to keep up with the rapidly changing phase screen seen by the telescope that results in an extended halo of speckles oriented along the direction that the wind is blowing (see, e.g., \cite{malesGroundbasedAdaptiveOptics2018} for an analysis on how this affects the raw contrast in the AO correction zone).

There are strategies that may be employed to combat the WDH. When imaging point sources, high-pass filtering or aggressive PSF subtraction may perhaps be employed with limited efficacy. However, spatial filtering is detrimental to diffuse, often faint structures associated with circumstellar disks (\cite{milliImpactAngularDifferential2012}, \cite{folletteComplexSpiralStructure2017}). One way to approach this phenomenon is with improvements to the AO control loop itself, either with faster loop speeds or higher AO loop gain. However, the AO loop speed is often limited by WFS detector noise or the capabilities of the realtime controller computer. It is also not trivial to choose the optimal gain coefficient for each controllable mode and overshooting this gain causes the AO loop to become unstable \cite{malesGroundbasedAdaptiveOptics2018}. There is value in mitigation strategies for the WDH, as simulations have shown improvements in raw contrasts on the order of $\sim1$ magnitude in the areas closest to the star (\cite{malesGroundbasedAdaptiveOptics2018}; \cite{cantalloubeWinddrivenHaloHighcontrast2020}).

Subtracting the point-spread function (PSF) is a challenging aspect of high-contrast imaging post-processing especially for resolved extended objects like circumstellar disks. This is because the true morphology of the extended object is typically not known beforehand, and thus an optimal PSF subtraction strategy that best preserves the signal of interest is also not known. An effective high-contrast imaging technique for circumstellar objects is Angular Differential Imaging (ADI; \cite{maroisAngularDifferentialImaging2006}). ADI involves fixing the telescope pupil with respect to the sky and letting the astrophysical scene naturally rotate through an observation which typically spans hours. It has been demonstrated that an improvement to classical ADI is Karhunen-Loéve Image Projection (KLIP; \cite{soummerDETECTIONCHARACTERIZATIONEXOPLANETS2012}, \cite{amaraPynPointImageProcessing2012}) as this approach excels at reconstructing and subtracting the majority of the PSF. However, the drawback to this approach is the unavoidable subtraction of disk (or planet) flux along with the noise (\cite{milliImpactAngularDifferential2012}; \cite{pueyoDETECTIONCHARACTERIZATIONEXOPLANETS2016}). Given this, it is critical to optimally choose the hyperparameters associated with \adi{} to best preserve the astrophysical signals of interest since there is currently no way to do this automatically. For disks, one strategy is to perform PSF subtraction with KLIP using the smallest number of KLIP modes in the basis as possible such that flux from diffuse structure survives the process. Going further, KLIP forward modeling (FM) methods are being developed with the goal to recovering the astrophysical signal lost to PSF subtraction (e.g., \cite{mazoyerDiskFMForwardModeling2020}; \cite{kuenyProbingDiskIce2024}; \cite{homUniformAnalysisDebris2024}) by simulating the instrument response and PSF subtraction artifacts on a model object. Reference Differential Imaging (RDI; \cite{ruaneReferenceStarDifferential2019}, \cite{xieReferencestarDifferentialImaging2022}) is an alternative method to ADI where images of another star that does not harbor an astrophysical object is used to subtract the PSF from the science target images. This imaging technique is better at preserving extended object features, but is also often limited by the WDH.

Techniques such as SCIDAR (Scintillation Detection and Ranging; \cite{verninExperimentalDeterminationTwodimensional1973}; \cite{roccaDetectionAtmosphericTurbulent1974}) and SLODAR (Slope Detection and Ranging; \cite{wilsonSLODARMeasuringOptical2002}) established the use of cross-correlation (CC) to ascertain the atmospheric wind profile and thus a quantifiable estimate on the observing conditions. Under the Taylor frozen-flow hypothesis \cite{taylorSpectrumTurbulence1938}, layers of turbulence present as translating static 2D intensity patterns in the telescope pupil (e.g., \cite{schockMethodQuantitativeInvestigation2000}). Given this idea, turbulence is often quantified using the atmospheric turbulence coherence time parameter, $\tau_0$, which is a direct measure of how fast the wavefront is changing due to the wind (units typically in ms). $\tau_0$ also dictates the AO loop speed required for ideal correction. Using Generalized SCIDAR, \cite{klueckersProfilingAtmosphericTurbulence1998} demonstrated one of the first on-sky retrievals of high-resolution wind velocity and refractive index structure constant $C_n^2$ profiles using a dedicated instrument. However, the wavefront sensor (WFS) is one of the main components within an AO instrument and offers a way to quantify turbulence within a telescope's line-of-sight without the need of a separate instrument. By applying the CC to WFS data using a range of time delays, the resulting response maps are images of moving peaks that define the speed and direction of discrete wind layers in the atmosphere; similar concepts were recently demonstrated using the Shack-Hartmann WFS on the Gemini Planet Imager instrument (\cite{levinsteinEstimatingEffectiveWind2022}, \cite{duEstimatingAtmosphericWind2024}).

There is value in a real-time description of the optical turbulence within the line-of-sight of an AO-assisted telescope (e.g., \cite{guesalagaUsingC2Wind2014}) as this has potential to supplement predictive control techniques that should drastically reduce the servo-lag error (\cite{poyneerFourierTransformWavefront2007}; \cite{malesGroundbasedAdaptiveOptics2018}). Additionally, turbulence nowcasts will lead to improved scheduling and instrument optimization which may save considerable on-sky time. Thus, profiling the turbulence above a site using instruments inherent to an AO system continues to be an active area of research. Recent work by \cite{dungeeCharacterizingAtmosphericTurbulence2022, mcewenWavefrontProfilingCorrelation2022} showed that cross-matching the wind vectors measured from the WFS telemetry with weather data led to increased knowledge on the turbulence above Mauna Kea. For the Cerro Paranal site, \cite{osbornAtmosphericTurbulenceForecasting2018} demonstrated a method to forecast the turbulence using a general circulation model from the European Centre for Median range Weather Forecasts with good agreement with stereo-SCIDAR measurements.

In this study, we detail our efforts towards WDH mitigation in coronagraphic science images of a circumstellar disk. To do this, we make use of pyramid WFS telemetry to profile the behavior of the atmospheric turbulence during the observations to place constraints on our models. In the following Section \ref{sec:observations} we describe our observations and data reduction techniques. In Section \ref{sec:pipeline}, we detail our implementation of a WFS data processing pipeline to retrieve the wind vectors of turbulent layers within the field of view of the telescope. We then follow with Section \ref{sec:retrieved-wind-vectors} where we report our pipeline output for the datasets used in our analysis. Section \ref{sec:forward-modeling} follows, where we describe our WDH forward modeling procedures and we report our resulting WDH model solutions in the proceeding Section \ref{sec:on-sky-demonstration}. We discuss our retrieved wind vectors and the implications of our WDH modeling solutions in Section \ref{sec:discussion-windsocc} and Section \ref{sec:discussion_windfm}, respectively. Finally, we summarize our analysis in Section \ref{sec:summary} and also include suggestions for future extensions to this project.

\section{Observations}
\label{sec:observations}

\begin{table*}[!ht]
\centering
\small
\caption{MagAO-X Observation Log of \hr{}}
\begin{tabular}{lcclccccccc}
\hline
Date        & Time         & Seeing$^a$  & Band  & Air Mass & $t_{\text{exp}}$ & $N_{\text{exp}}$ & $N_{\text{coadds}}$ & $\theta$ & $\tau_0^b$  \\
            & (UT)         & $(")$       &       &          & (s)              &                  &                     &  $(^{\circ})$ &  (ms) \\ \hline
2023 Mar 10 & 05:49/07:36 & 0.6 &  $i'/z'$  & 1.0/1.1  & 1                    & 5047    &     84         & -21.5/65              &    6.4    \\
2023 Mar 13 & 04:46/07:15 & 0.6 & $r'/g'$  & 1.1/1.0  & 0.23                 & 35355   &   122               & -58.5/61.5             &     5.0    \\

\hline
            &              &            &       &          &                      &                     &                  &                   &   &               
\end{tabular}

\footnotesize
\raggedright
\textbf{Notes.} Time: start/end observation; Air Mass: start/end air mass; $t_{\text{exp}}$: single frame exposure time; $N_{\text{exps}}$: number of raw science images; $N_{\text{coadds}}$: number of image cutouts used for PSF subtraction;  $\theta$: start/end parallactic angle. The total integration time for a given dataset is found by multiplying $t_{\text{exp}}$ and $N_{\text{exp}}$. \\
\vspace{0.5mm}
$^a$ median values in arcseconds.
$^b$ calculated using the measured jet stream velocities with Eq. 5 from \cite{sarazinESODifferentialImage1990} to obtain an estimate of the effective wind velocity; see Section \ref{sec:discuss-compare-era5-subsec}.
\label{tab:obslog}
\end{table*}

We observed \hr{} on 2 separate nights using the 6.5 m Magellan-Clay telescope at Las Campanas Observatory (LCO) in Chile. These observations took place during the 2023A semester at \gband{} (527 nm, effective bandwidth $\omega_{\text{eff}} = 41$ nm), \rband{} (615 nm, $\omega_{\text{eff}} = 110$ nm), \iband{} (762 nm, $\omega_{\text{eff}} = 126$ nm), and \zband{} (908 nm, $\omega_{\text{eff}} = 132$ nm). On both of these nights, we observed \hr{} in dual-imaging mode using either the $i'/z'$ or $r'/g'$ filter configuration; see Table \ref{tab:obslog}. MagAO-X operates with the pupil fixed with respect to the sky to facilitate ADI. Additionally, we made use of MagAO-X's small Lyot coronagraph (focal plane mask radius of about 21 mas or $\lesssim3 \lambda/D$) for all observations.  The instrument has a plate scale of 5.9 mas/pixel \cite{longAstrometricCalibrationMagAOX2025} and a field-of-view of $6\farcsec0 \times 6\farcsec0$. The high-order AO loop speed was set to 2 kHz for the observations presented in this paper.

MagAO-X has a suite of 3 deformable mirrors (DM) with the 2k-actuator DM operating in the high-order AO loop acting as the workhorse of the instrument. Using this DM, we can generate four incoherent calibration DM speckles (\cite{mcewenOnskyRealtimeOptical2024}) by setting a sinusoidal pattern on the reflective face sheet that are useful for, e.g., astrometry (to locate the star behind the coronagraph), photometry, and distinct image artifacts for frame registration \cite{jovanovicArtificialIncoherentSpeckles2015}. These artificial speckles offer a convenient way to measure the Strehl ratio of the PSF during coronagraphic observations and so we used these image features to exclude frames of poor quality (explained in Section \ref{sec:forward-modeling}). For our \hr{} observations, we generated speckles at a distance of $15 \lambda/D$ from the star (the AO control radius goes out to 24 $\lambda/D$).

\section{\texttt{windsoCC} pipeline description}
\label{sec:pipeline}

\begin{figure}[ht]
\includegraphics[width=\linewidth]{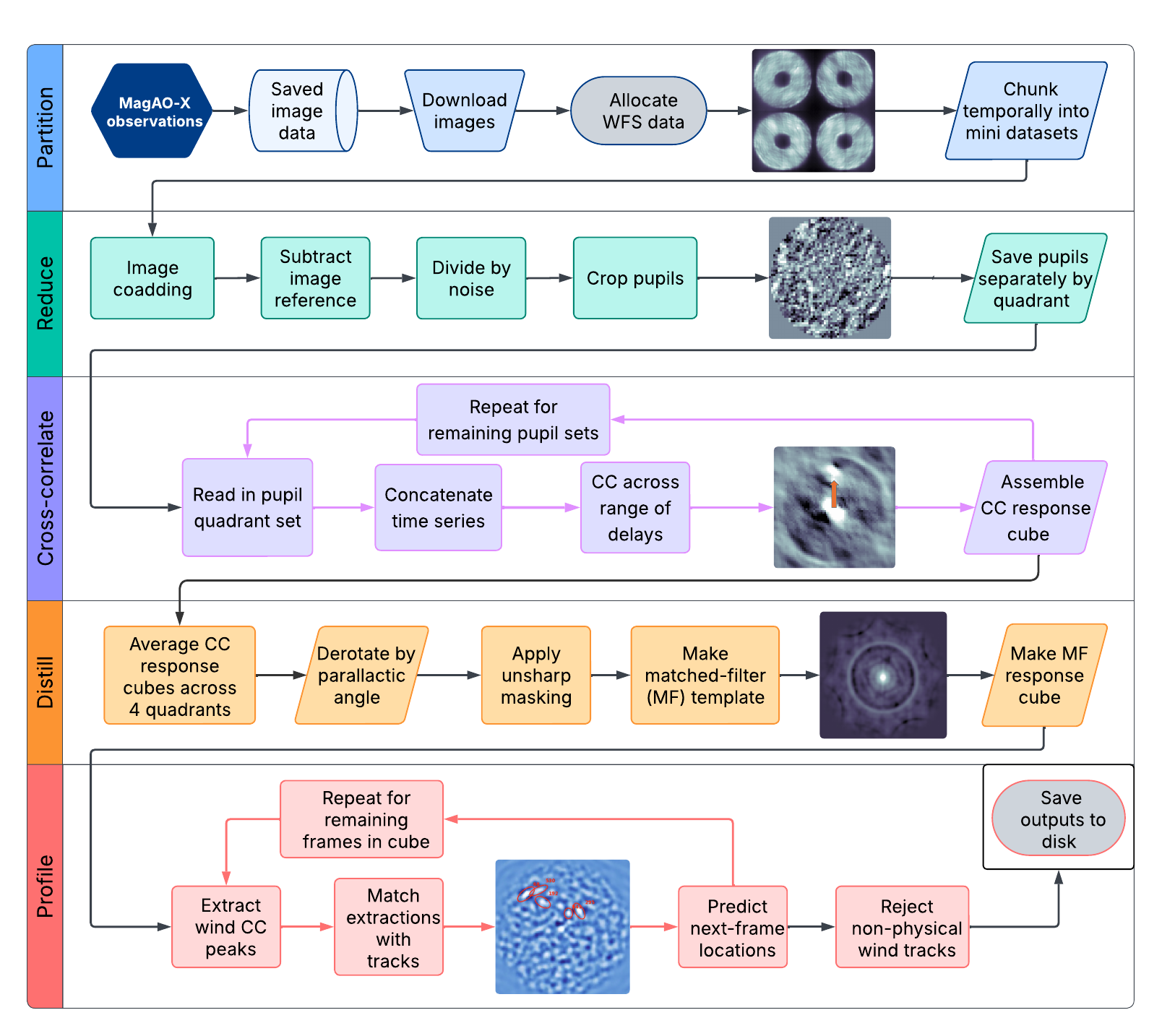}
\caption
{Flowchart illustrating the various operations performed in the \texttt{windsocc} pipeline in left to right progression. This workflow is explained in detail in Section \ref{sec:pipeline}. We inset small thumbnail images illustrating data products at the intermediate steps. The final outputs include a human-readable text file that contains the information on each detected wind layer and the MF response cubes. We adhere to specific flowchart shapes to denote the following: rectangles for processing steps, trapezoids for manual operations, parallelograms for intermediate data products that are saved to disk, hexagons for preparatory steps, cylinders for hard disk storage, and gray pill shapes for the pipeline's starting/ending operation.}
\label{fig:flowchart}
\end{figure}

In order to gauge how different layers of atmospheric turbulence affect MagAO-X's performance, we implemented a pyramid WFS \cite{ragazzoniPYRAMIDWFS1996} image processing pipeline, \texttt{windsoCC}, based on CC. The basic premise is that under the assumption of frozen-flow, turbulent layers will appear as 2D static phase screens undergoing translation across the telescope pupil images. Then, by computing the CC between a pupil image with itself at some later time, the CC response will be a distinct peak at some distance from the origin. The displacement of the CC peak as a function of temporal delay therefore provides a direct observable for wind vectors projected onto the pupil image during data acquisition. Since the CC peaks encode the residual error that propagates to the science focal plane, we can then leverage the \texttt{windsoCC} output to inform our WDH model parameter priors when reconstructing the WDHs. Our pipeline follows a number of discrete stages that we illustrate using a flowchart in Figure \ref{fig:flowchart}. We follow with descriptions of each pipeline stage in the next few subsections.

\subsection{Partition: re-organizing the raw telemetry}
\label{sec:partition-subsec}

The MagAO-X modulated pyramid WFS records telemetry typically at a cadence of $\geq 2000$ Hz (the AO loop speed was 2 kHz for the observations presented in this paper), producing a large number of image data FITS files over the course of a science observation. For characterizing the wind, these data must be grouped on timescales short enough that the behavior of the atmospheric turbulence can be treated as approximately stationary, while also spanning sufficient integration time to minimize the noise floor. We therefore partition the raw WFS telemetry into short, time-ordered segments before running the subsequent reduction, cross-correlation, and wind-measurement stages.

The partitioning step identifies raw \texttt{camwfs} FITS cubes, sorts the files chronologically, and groups them either into fixed-duration intervals, typically on the order of a few seconds, or into a specified number of equal groups. Each group is moved into a dedicated subdirectory and treated as a separate dataset during the later stages in the pipeline. This organization defines the period of time over which wind properties are measured and later quantified. For the analysis presented in this manuscript, we elected to partition the dataset into groups encompassing 12 seconds of WFS integration time (in this study, since we binned the raw frames temporally in 8 frame chunks, this corresponded to 3k frames) as we found this led to a good balance between reducing noise and sampling rate. In this way, we probe for wind vector information on the order of 5 times per minute across the observation.

\subsection{Reduce: WFS image preparation and calibration}
\label{sec:reduce-subsec}

After partitioning the raw telemetry into shorter temporal segments, each set of pyramid WFS images is calibrated and reformatted for cross-correlation analysis. The purpose of this reduction step is to remove quasi-static detector and illumination structure from the raw WFS frames while preserving just the residual wavefront structure. This pre-processing is critical since the subsequent CC stage is sensitive to static aperture features and detector artifacts and makes the CC response ambiguous.

For each partitioned telemetry directory, \texttt{windsoCC} first identifies the available \texttt{camwfs} FITS cubes and loads the corresponding reduction parameters from the pipeline configuration file. The pipeline constructs a reference image by combining all FITS cubes in the working directory by averaging (though this could also be done with a median stacked image). The second calibration product the pipeline generates is an estimate of the spatial noise by computing the per-pixel standard deviation across all FITS cubes in the working directory. The reference image represents the mean pupil illumination pattern, which is subtracted off each frame, while the noise map is used to normalize the data. Each frame is then cropped around the 4 pyramid WFS pupil images. The output of this stage is a set of reduced images for each WFS pupil quadrant, saved separately for the upper-left, upper-right, lower-left, and lower-right pupils. These reduced pupil cubes form the calibrated input data for the cross-correlation analysis used to recover wind velocities.

The 4 pupil images in each quadrant of the \texttt{camwfs} frames are bisected by amplifier boundaries on the EMCCD (electron-multiplied charge-coupled device) chip which caused a strong high-frequency pattern artifact in the CC response maps along the vertical direction. For the datasets used in this manuscript, we found that coadding in 8 frame chunks was sufficient for averaging this artifact out. Even with this binning operation, since the AO loop speed during acquisition of these data was 2kHz (single frame integration time of 0.5 ms), we still maintained the sampling rate necessary to resolve atmospheric motion much exceeding the expected jet stream speeds (typically 20-50 m/s). As a concrete example, a wind layer traveling at 150 m/s would travel about 0.6 meters (or about a tenth of the diameter of the Magellan-Clay primary mirror) in 4 ms (i.e., 8 WFS frames at a 2 kHz AO loop speed) and thus would be easily trackable in our data.

\subsection{Cross-correlate: retrieving the turbulence layers}
\label{sec:xcorr-subsec}

The reduced WFS pupil cubes are analyzed using the CC to measure the apparent motion of residual wavefront structure across the pupil. For each partitioned data segment, \texttt{windsoCC} processes the reduced FITS cubes from each pyramid WFS pupil quadrant independently. To maximize the signal-to-noise of the CC peaks due to turbulent wind, \texttt{windsoCC} uses a segmented, overlapping CC method, conceptually analogous to the variance-reduction techniques used in Welch's method for power spectral density estimation \cite{welchUseFastFourier1967}.

Recall that \texttt{windsoCC} works with the miniaturized 12-second datasets produced during the partition stage. Within each 12-second block, a maximum correlation time delay ($\Delta t_{max}$) of 2 seconds was used for this dataset to capture the slow-moving ground layers (though this delay is configurable). The reduced frames in each block are subdivided into a series of overlapping segments of duration $\Delta t_{max}$. To ensure uniform sampling and recover signal near the segment boundaries, our CC method uses a 50\% overlap between adjacent segments. For example, the first segment comprised the frames spanning $t = 0$ to $2$ seconds, the second segment spanned $t = 1$ to $3$ seconds, the third $t = 2$ to $4$ seconds, and so forth. Then, the CC is performed over a configurable range of frame delays.

The CC was computed independently for each 2-second segment, generating a localized response cube. Finally, the response cubes for all overlapping segments within the 12-second block were temporally averaged. This ensemble averaging significantly reduces the variance of the CC response background, suppressing noise while reinforcing the coherent phase residuals of strong wind layers.

\subsection{Distill: enhancing CC response maps}
\label{sec:distill-subsec}

\begin{figure}
\begin{center}
\begin{tabular}{c}
\includegraphics[width=\linewidth]{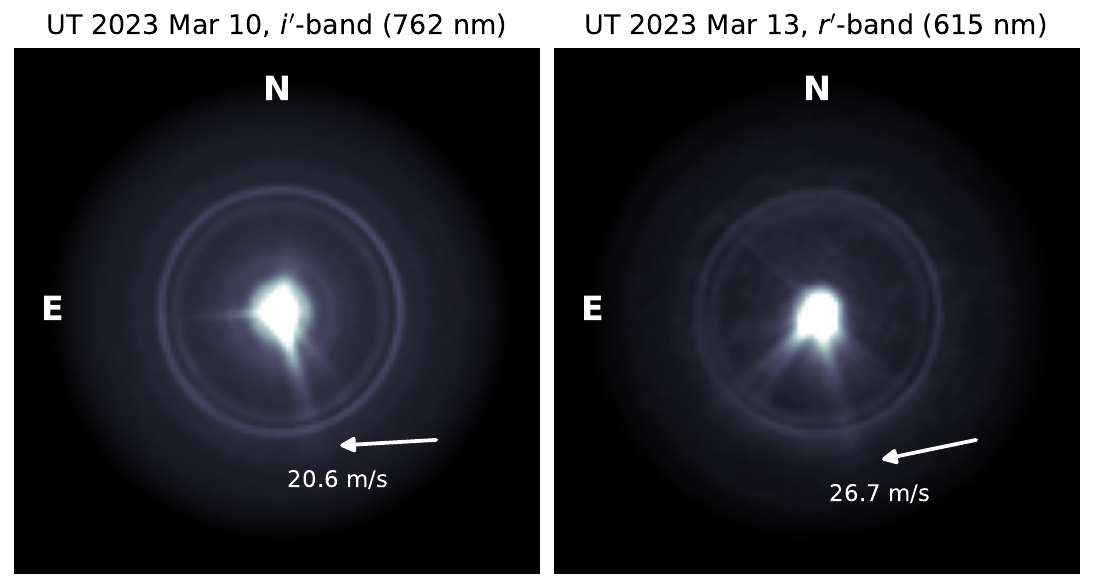}
\end{tabular}
\end{center}
\caption
{Time-collapsed CC response maps illustrating the dominant layers of turbulence and their directions through the observation. We used a maximum delay of 2 seconds for the CC. The outer ring-like artifact corresponds to the edge of the primary mirror edge in the WFS data. For reference, we illustrate the measured mean speed and direction of the fastest layer detected during the observation as the white arrow in the bottom right. To confirm that the fastest moving peaks correspond to the jet layer, we cross-matched our measurements with historical weather data (ERA5).}
\label{fig:wind-collpased}
\end{figure}

The cross-correlation stage produces one response cube for each of the four pyramid WFS pupil images. Because each quadrant samples the same wavefront but through different Fourier modes \cite{guyonLimitsAdaptiveOptics2005}, \texttt{windsoCC} combines these independent measurements to increase the signal-to-noise of the CC response maps. For each partitioned time segment, the pipeline groups the four quadrant-specific cross-correlation cubes by their common timestamp and averages them into a single combined response cube. Next, \texttt{windsoCC} accounts for the changing orientation of the pupil relative to the sky by derotating by the corresponding parallactic angle so that the wind response maps are aligned with the sky across the observation.

The distillation stage also produces several diagnostic and analysis products. A version of the combined response cube is subjected to unsharp masking (to suppress low-spatial frequency structure while preserving the correlation peaks due to the wind. A matched-filter (MF) template is constructed from the central region in the zeroth delay frame (i.e., the autocorrelation of the residual phase pattern in the aperture) and applied to the sharpened frames. The resulting MF response cubes provide enhanced signal-to-noise for the CC peaks which helps later source detection and wind tracking. For an interesting visualization of the movement of the CC peaks across the observation, we combined the CC response cubes for each night and then collapsed them into a single image (Figure \ref{fig:wind-collpased}). This serves as a quick way to gauge the directions of the most turbulent atmospheric layers for a given observation. We include the direction vector of the fast jet stream layer as the white arrow.

\subsection{Profile: tracking and measurement of wind}
\label{sec:tracking-and-measurement-of-wind-subsec}

Wind layer attributes are measured from the MF response cubes produced by the distillation stage. In these cubes, a wind layer appears as a compact peak whose displacement from the origin increases linearly and with constant velocity with temporal delay. The measurement stage therefore detects and tracks coherent peaks due to individual wind layers as they move through the correlation space. Early delay frames are ignored because physically meaningful wind peaks have not yet propagated far enough from the origin (i.e., positive detections from these frames would log excessively high wind speeds of $\gtrsim100$ m/s). In subsequent frames, candidate peaks are detected with Source Extractor facilitated by the \texttt{SEP} Python library (\cite{bertinSExtractorSoftwareSource1996a}; \cite{barbarySEPSourceExtractor2016}).

\begin{figure}
\begin{center}
\begin{tabular}{c}
\includegraphics[width=\linewidth]{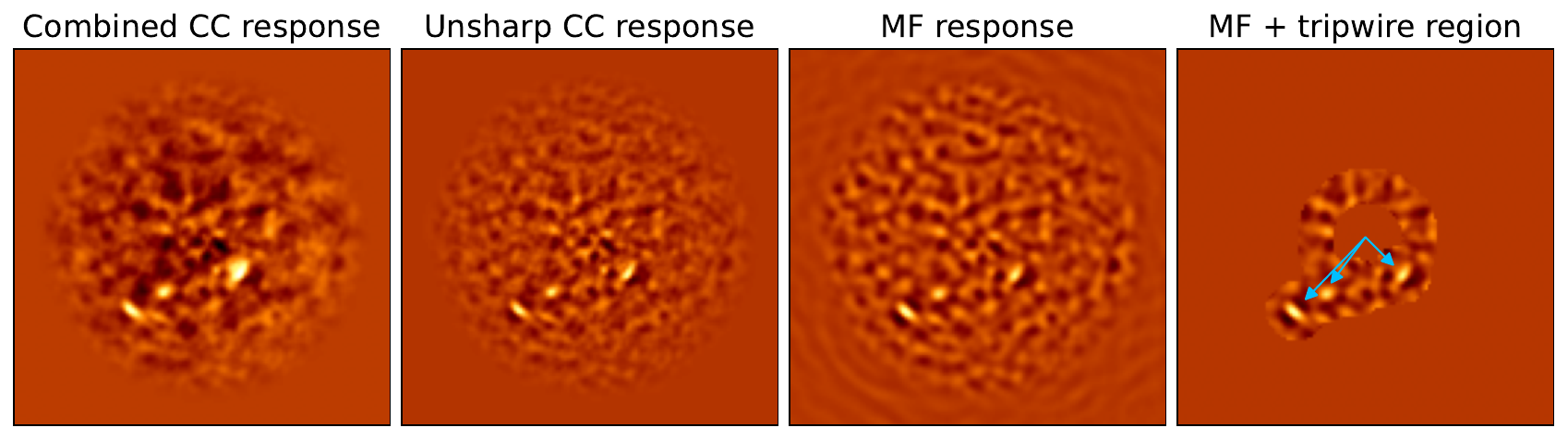}
\end{tabular}
\end{center}
\caption
{Visuals showing the pipeline processes applied to the CC response maps using a representative frame in a cube. All images are plotted using the same scaling. \textbf{Left:} CC response map after combining the CC responses from each pupil quadrant (``distill" pipeline stage). \textbf{Left middle:} combined CC response after applying unsharp masking. \textbf{Right middle:} response map after performing a MF operation after unsharp masking. \textbf{Right:} the MF response frame shown with an example ``tripwire" search region spanning 12 to 24 pixels from the center of the image. The radially-outward motion of the 3 CC peaks visible in these frames is marked with blue arrows on the rightmost image. For each frame in the cube, the tripwire search region is scanned for peaks, and detections that are consistently matched with a track are spotlighted with a dynamic circular window until it is either dropped due to low signal-to-noise or it reaches the edge of the frame. We show an example dynamic circular window plopped on the furthest peak in the rightmost image.}
\label{fig:tracking-wind-demo}
\end{figure}

For each high-pass matched-filter response cube, \texttt{windsoCC} defines an annular ``tripwire'' region around the central autocorrelation peak whose width is configurable where \texttt{SEP} will search for peaks and ignore the rest of the image (Figure \ref{fig:tracking-wind-demo}). The concept of the tripwire search region was motivated by the idea that all peaks due to a wind layer will traverse this region going (1) linearly, (2) radially-outward, and (3) at a constant velocity. Peak detections that are due to noise will not follow these 3 rules and thus become easy to dynamically reject. For these data, we set the tripwire region to span the region between 12 and 24 pixels from the autocorrelation peak (see Figure \ref{fig:flowchart}, ``Distill" lane for an inset response map with the central autocorrelation peak). We measured the full-width at half-maximum value of the autocorrelation peak to be $\sim1.5$ pixels, so the width of our tripwire region was sufficiently-wide such that \texttt{SEP} would have multiple opportunities to recognize a peak. Importantly, reducing the area of the search region for \texttt{SEP} drastically improved the computation time during this step and streamlined the rejection of garbage detections.

The next step after using \texttt{SEP} to extract the sources in a given frame is to then match these candidate sources to a wind track. Linking source detections to a track is done via the Hungarian method \citep{kuhnHungarianMethodAssignment1955} using distance from the current frame's expected location for that track. As an additional guard against garbage detections, a wind-tracking model vets candidate sources using the expected continuity of peak position, direction, and speed. Dynamic masking is used to reopen regions around previously tracked candidates, allowing wind peaks to continue being followed after crossing the initial tripwire region (as illustrated in Figure \ref{fig:tracking-wind-demo}, rightmost panel).

Each accepted track is converted into a wind measurement using the known plate scale of the WFS pupil and the temporal sampling of the reduced data. The spatial displacement of a track gives the projected direction of motion in the WFS frame, while the change in displacement with delay gives the wind speed in, e.g., meters per second. Finally, all individual detections within each track are examined as an ensemble and tracks are filtered using criteria on the total number of detections, the inferred track origin, linearity of motion (Rule 1), outward propagation (Rule 2), and scatter in direction and velocity (Rule 3). The surviving tracks are written as per-cube wind vector files.

\subsection{Quantifying wind attributes}
\label{sec:quantifying-wind-attributes-subsec}

The final \texttt{windsoCC} measurement products summarize each accepted wind track with a small set of physically interpretable parameters. For every matched-filter response cube, the pipeline reports the wind layer track identifier, wind speed, propagation direction, and the peak's average area in pixels. The resulting files form a time-resolved lookup table of exceptionally-turbulent wind layers during the observation.

\section{\texttt{windsoCC} Results: retrieved wind vectors}
\label{sec:retrieved-wind-vectors}

\begin{figure}
\begin{center}
\includegraphics[width=0.48\linewidth]{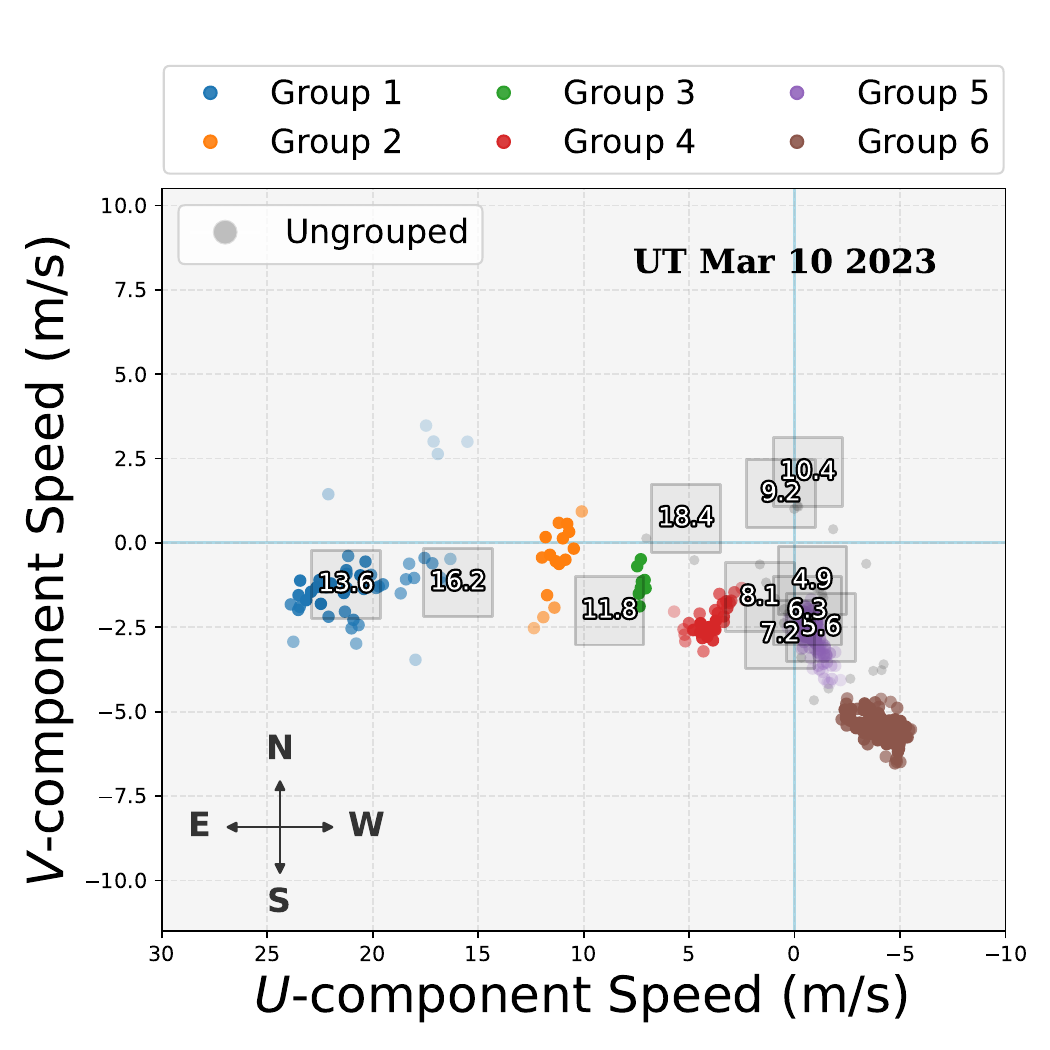}
\includegraphics[width=0.48\linewidth]{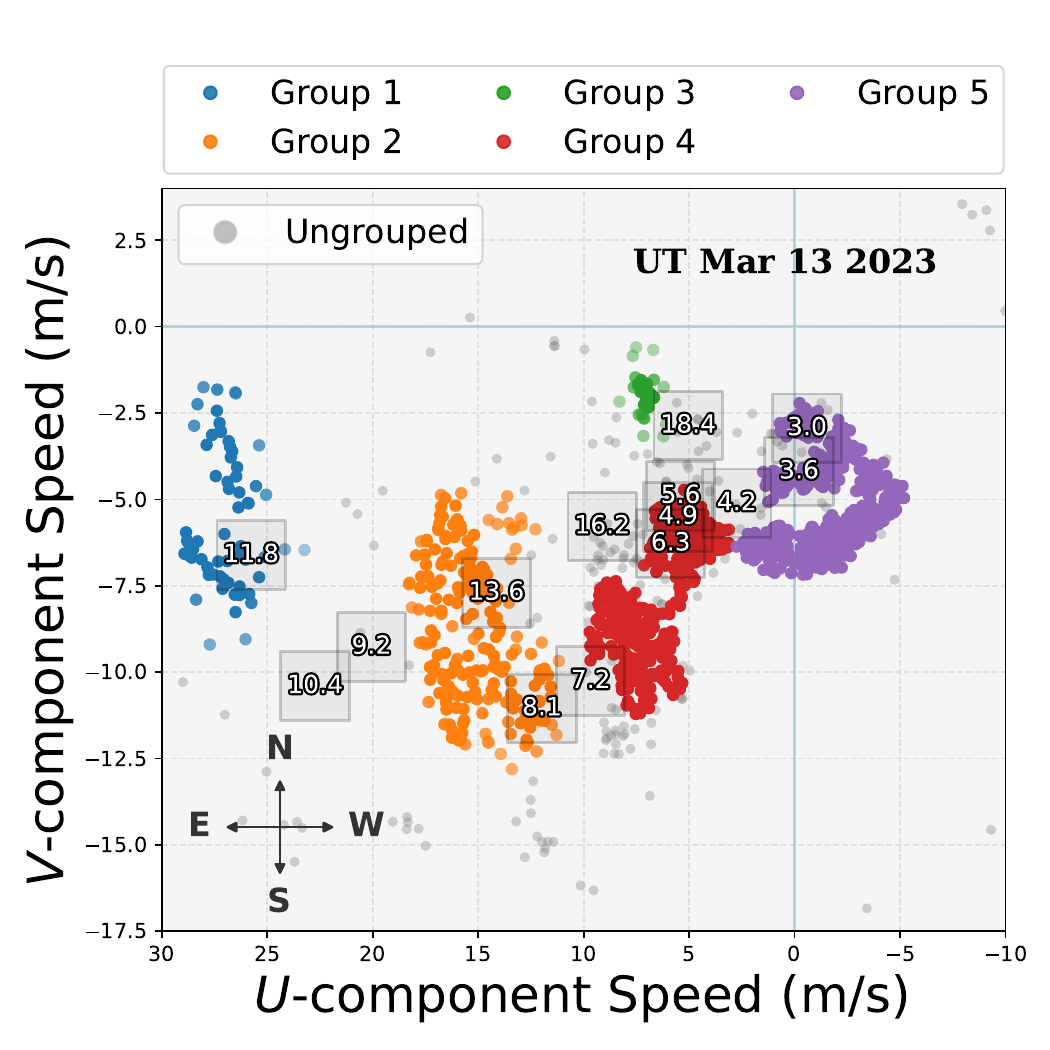}
\end{center}
\caption{Clustering analysis on wind layer velocities and directions extracted by the \texttt{windsocc} pipeline. The vertical and horizontal axes show the V-component ($v\sin\theta$) and U-component ($v\cos\theta$) of the wind direction vectors respectively. We used the Hierarchical Density-Based Spatial Clustering of Applications with Noise (HBDSCAN) algorithm to facilitate cluster identification in this feature space. We assign colors to the identified groups according to the legend on top of the plots and the relative opacity of the assigned colors represents the probability of individual points belonging to that colored cluster. We illustrate points identified as noise by HBDSCAN as light gray dots. For reference, we plot the altitudes in kilometers of wind layers using the ERA5 wind data contained in Table \ref{tab:era5} as the numbers in the light gray squares.}
\label{fig:clusters-wind}
\end{figure}

\begin{table*}[h]
\centering
\footnotesize
\caption{\texttt{windsoCC} extracted wind vectors.}
\vspace{6pt}
\hspace*{-2cm}
\begin{tabular}{ccc|cc}
\hline \hline
 & \multicolumn{2}{c}{\textbf{UT 2023 Mar 10}} & \multicolumn{2}{c}{\textbf{UT 2023 Mar 13}}  \\ \cline{2-5} 
                  & Mean Velocity        & Mean Direction     & Mean Velocity      &   Mean Direction   \\ 
                  & [m/s]                  & $[^\circ]$            & [m/s]                &   [$^\circ$]   \\ \hline
Group 1           & 20.64 $\pm$ 2.19     & 93.32 $\pm$ 3.47   & 27.51 $\pm$ 1.18   & 101.72 $\pm$ 4.04  \\
Group 2           & 11.28 $\pm$ 0.64     & 92.57 $\pm$ 4.90   & 17.64 $\pm$ 1.30   & 120.28 $\pm$ 7.35  \\
Group 3           & 7.37 $\pm$ 0.14      & 99.25 $\pm$ 3.17   & 7.38 $\pm$ 0.42    & 106.65 $\pm$ 3.98  \\
Group 4           & 4.59 $\pm$ 0.87      & 119.95 $\pm$ 3.83  & 9.63 $\pm$ 2.05    & 140.09 $\pm$ 5.17  \\
Group 5           & 2.79 $\pm$ 0.55      & 196.98 $\pm$ 6.76  & 5.50 $\pm$ 1.21    & 194.83 $\pm$ 17.11  \\
Group 6           & 6.87 $\pm$ 0.66      & 217.17 $\pm$ 4.89  & ---                & ---                \\
All               & 8.92 $\pm$ 2.59      & 136.54 $\pm$ 11.42 & 13.53 $\pm$ 2.98   & 132.71 $\pm$ 20.14 \\ \hline
\end{tabular}

\footnotesize
\vspace{2mm}
\raggedright \textbf{Notes.} 1 sigma uncertainties associated with each cluster of points.

\label{tab:wind-stats}

\end{table*}

\begin{figure}
\begin{center}
\includegraphics[width=0.45\linewidth]{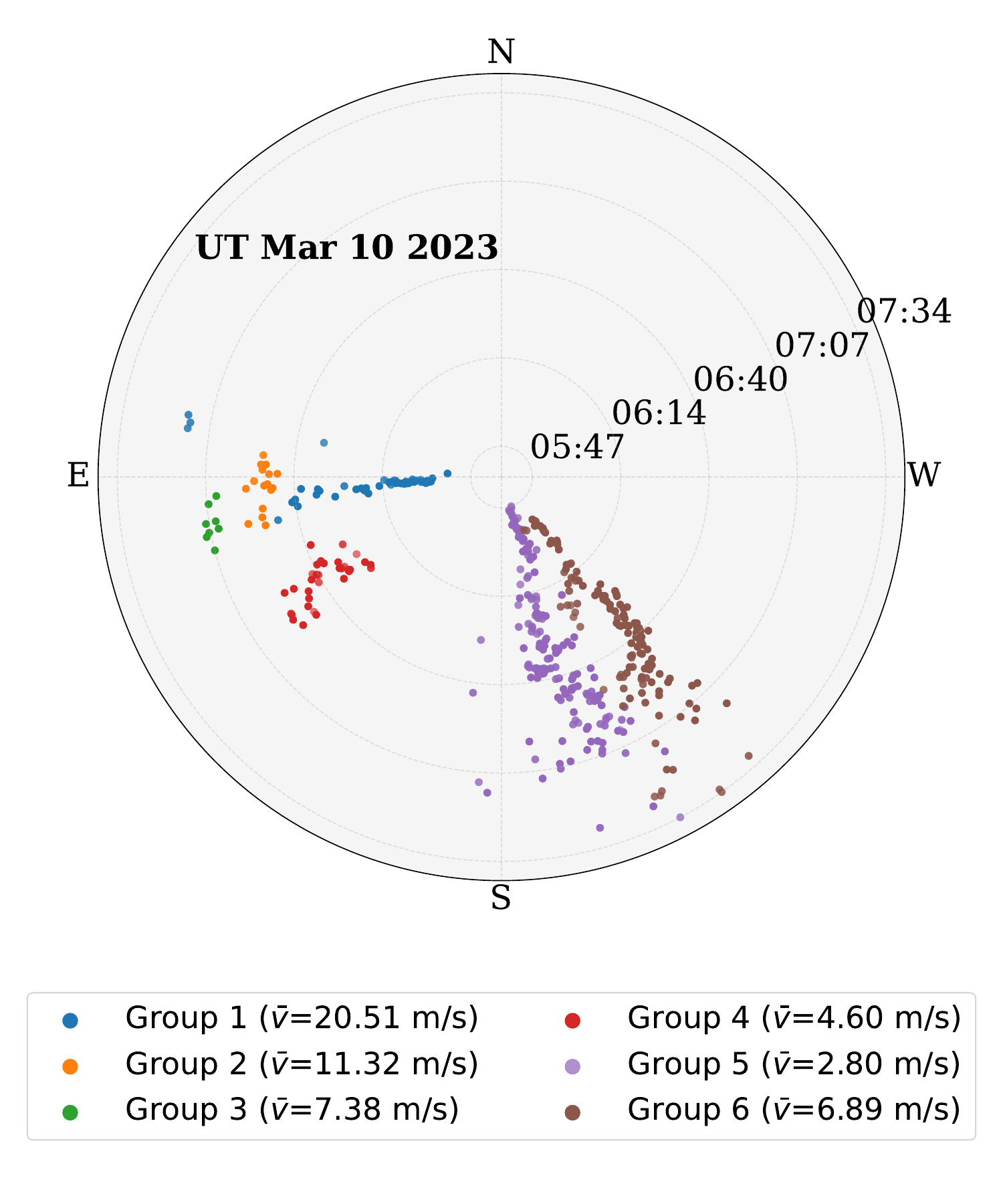}
\includegraphics[width=0.45\linewidth]{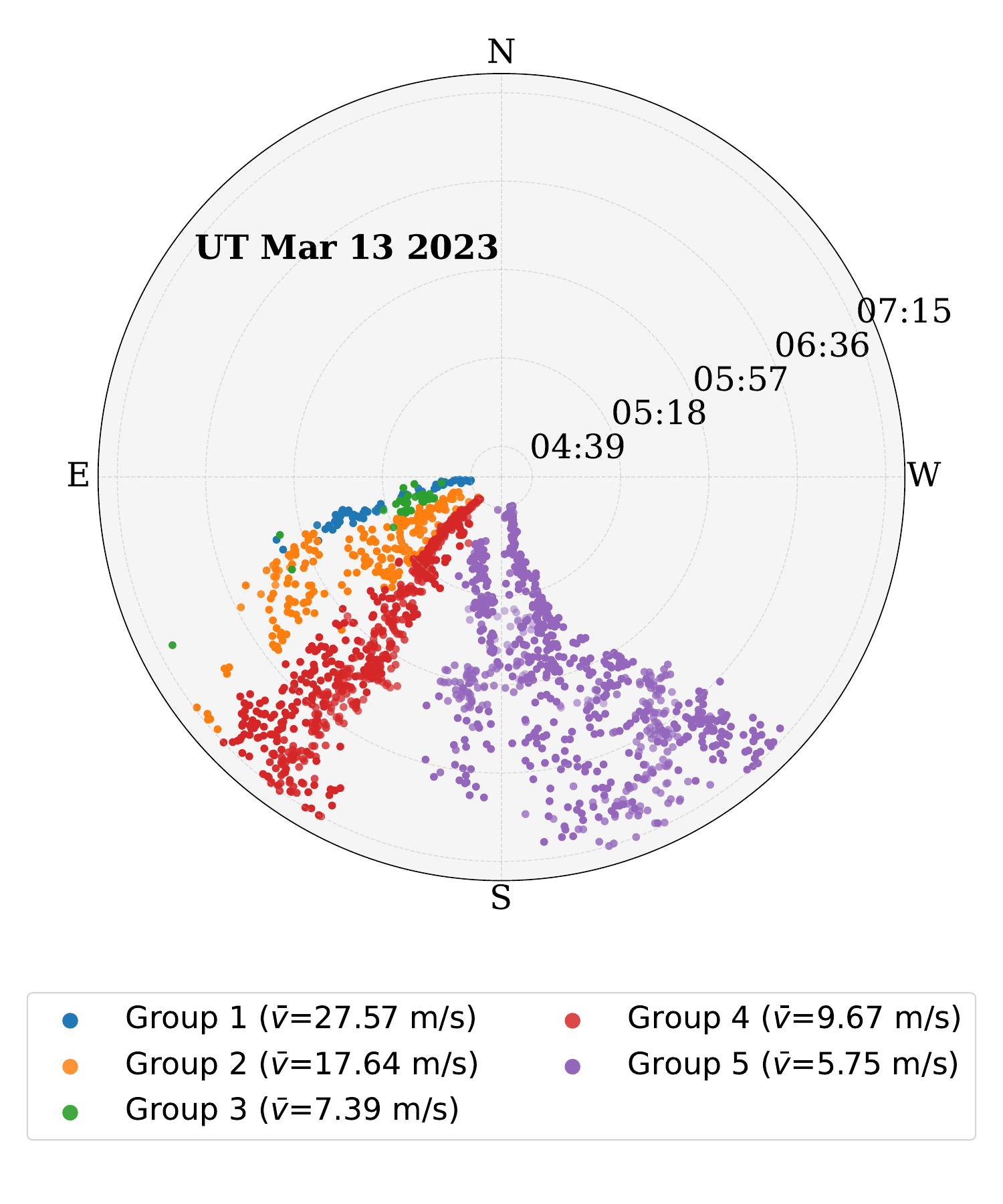}
\end{center}
\caption{Plotting the wind points belonging to each cluster in Figure \ref{fig:clusters-wind} across time for the datasets from UT 2023 Mar 10 (left) and UT 2023 Mar 13 (right). We show the time in UT along the radial direction of the chart and the directions the wind layers are traveling in a North up, East left configuration.}
\label{fig:directions-vs-time-wind}
\end{figure}

We report the measured wind speeds and directions throughout the datasets from UT 2023 Mar 10 and 13 as a clustering visualization in Figure \ref{fig:clusters-wind} and tabulated in Table \ref{tab:wind-stats} with 1 sigma uncertainties. We also include a visualization of the wind direction versus time for each numbered cluster in Figure \ref{fig:directions-vs-time-wind}. In these figures, we use a North up, East left coordinate convention and the plotted points indicate the direction that the wind is blowing towards. This differs from the standard meteorological convention, but we felt that portraying the wind vectors in this manner facilitates comparison with the on-sky orientation of the WDH (see Section \ref{sec:forward-modeling}) and naturally follows the movement of the peaks in our CC response maps (Figure \ref{fig:wind-collpased}). For the clustering analysis, we made use of the Hierarchical Density-Based Spatial Clustering of Applications with Noise (HBDSCAN) algorithm as implemented by the \texttt{scikit-learn} machine learning library for Python. We used a minimum cluster size of 10 as well as a value of 0.5 for the distance threshold parameter for identifying clusters in the feature space shown in Figure \ref{fig:clusters-wind} (see Section \ref{sec:discuss-compare-era5-subsec} for our reasoning for these choices for the parameter values). In addition to assigning labels to points meeting a certain density threshold in the feature space, HBDSCAN also identifies sparse, unassociated points and assigns probabilities that grouped points belong to a certain cluster. We illustrate these sparse points as gray dots and cluster association probabilities as the relative translucence of the dots (completely opaque means 100\% probability the point belongs to the assigned group). For Figure \ref{fig:directions-vs-time-wind}, we show measured wind speeds slower than the ensemble maximum with dots that are more translucent.

\section{Forward modeling the wind-driven halo}
\label{sec:forward-modeling}

\begin{figure}
\begin{center}
\begin{tabular}{c}
\includegraphics[width=0.95\linewidth]{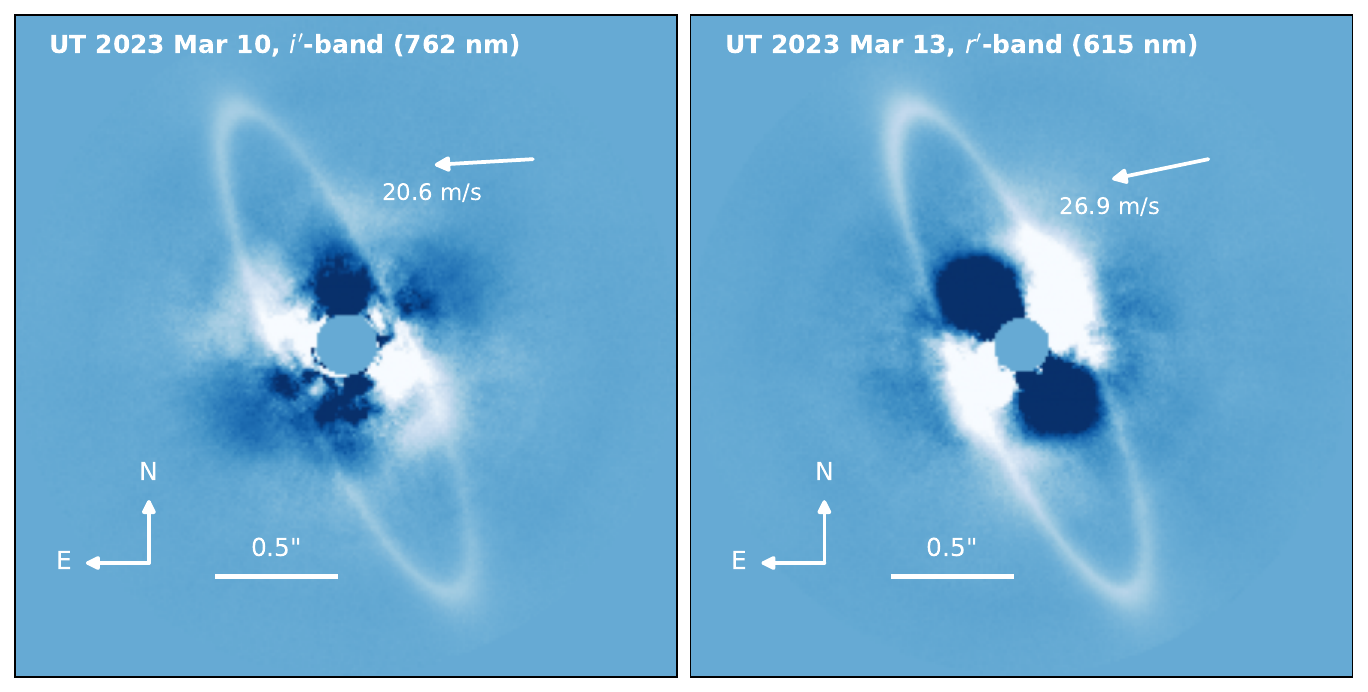}
\end{tabular}
\end{center}
\caption
{Example of the effect of the WDH on MagAO-X images reduced using \adi{}. Left: reduced image of \hr{} at \iband{} using data collected on UT 2023 Mar 10. Right: reduced \rband{} image using data collected on UT 2023 Mar 13. We subtracted the PSF from both images using 1 KLIP mode. Both images show the wind vector corresponding to the jet stream as the white arrow.}
\label{fig:wdh-klip}
\end{figure}

The WDH is both a high-contrast \cite{cantalloubeWinddrivenHaloHighcontrast2020} and dynamic feature of the PSF, making it challenging to ascertain its true morphology in the science images. This is made especially complicated given that the WDH may be caused by multiple fast layers of turbulence. To retrieve the best estimate of the morphology of the WDH in individual science frames, we elected to KLIP-reduce our images of \hr{} using conservative parameters (Figure \ref{fig:wdh-klip}) and then forward model a parametric model of the WDH in a Markov Chain Monte-Carlo (MCMC) framework. By analyzing the reduced image, we detect the artifact at high signal-to-noise and can determine how the WDH is surviving the current level of post-processing. Further, since we fed these data through our \texttt{windsocc} pipeline, we used the measured wind attributes to inform our parameter priors (such as the number of WDH models to use and their expected position angles) before running the MCMC.

For the dataset collected on UT 2023 Mar 10, we decided to fit 4 WDH models during the MCMC motivated by the wind layers detected by \texttt{windsoCC} (Figure \ref{fig:clusters-wind} and Figure \ref{fig:directions-vs-time-wind}, left panels). The groups that we fit WDH models to include group 1, 4, 5, and 6. Though group 1, 2, and 3 appear as distinct clusters in Figure \ref{fig:clusters-wind}, we elected to only fit one WDH model given that these wind vectors all point in roughly the same direction as shown in Figure \ref{fig:directions-vs-time-wind}. Namely, we hypothesized that the WDH artifact contributed to the final KLIP-reduced image (Figure \ref{fig:wdh-klip}, left side) by groups 1, 2, and 3 could be captured with a single WDH model component. 

We fit 5 models for the UT 2023 Mar 13. Though there were only 4 clusters identified by HBDSCAN that occupied distinct sections in direction space (Figure \ref{fig:clusters-wind}, right panel), we were inclined to fit 2 models to group 5 due to its large spread in wind direction. Indeed, visual inspection of the wind vectors recorded over time revealed that group 5 points are spread across a large directional range (Figure \ref{fig:directions-vs-time-wind}). It is possible that group 5 actually encompasses more than 2 layers of turbulence, but to avoid parameter bloat during our MCMC runs we ultimately decided to attempt to fit our WDH artifact in the reduced image using 5 individual models.

\subsection{Focal plane data reduction}
\label{sec:data-reduction}

To process the raw coronagraphic images for each of our datasets, we implemented the procedure described in \cite{kuenyMultibandStudyHR2026}. In short, we performed aperture photometry on the abovementioned artificial DM speckles to reject frames whose counts fell significantly below the ensemble average. Our exclusions criteria surrounding our DM speckle photometry resulted in excluding $7.5\%$ and $4\%$ of the raw frames for the \iband{} and \rband{} datasets respectively. We then made $448\times448$ image cutouts coarsely centered on the star and performed a phase cross-correlation technique to register the images to the sub-pixel level. Our next step entailed subtracting off the median radial profile off each individual image to remove the static stellar halo component. We then binned our registered image cutouts using $2\times2$ pixel binning and then coadded (i.e., binned temporally) using 60 second temporal bins. Following this, we performed a quick visual inspection of our registered frame cutouts by computing the median stack of the registered image cutouts and subtracting it from each of the cutouts. During this quick check we noticed artifacts in the AO control region, likely due to some modal gain values being higher than what was optimal given the observing conditions, which introduced loop instability; see \cite{poyneerOptimalModalFouriertransform2005}). These ``over-driving" artifacts were present in the last 44 minutes of the dataset from UT 2023 Mar 10 and in the first 50 minutes of the dataset from UT 2023 Mar 13. These artifacts typically manifest as characteristic bright blobs along the edge of the control region or surrounding the artificial DM speckles. We decided to exclude the frames that exhibited signs of over-driving to prevent these features from biasing our WDH model fits, as they are extended features and mirrored across the central star within the AO control region.

\subsection{Analytical model description}
\label{sec:model-description-subsec}

We developed an analytical model for simulating the WDH that assumes a Gaussian envelope along the direction of the wind and a ``pinch" parameter that dictates the opening angle of the wings. Our model also makes use of a flaring parameter, $\gamma$, where larger values causes the lobes to flare with radial distance (like the bell of a trumpet) and smaller values make for an hourglass shape. During some preliminary WDH modeling trials, we observed qualitatively-good fits using a 6-parameter model when comparing to example MagAO-X images where the WDH is visible in coronagraphic images as well as the example displayed in \cite{cantalloubeWinddrivenHaloHighcontrast2020}. To generate our parametric models, we use a 2D coordinate system $(x,y)$ representing position on the science camera detector and model the intensity $I(x,y)$ of the WDH captured at the focal plane using:

\begin{equation}
    I(x,y) = N_0 \exp{\left[-\frac{1}{2}\left(\frac{r}{p_0^{\gamma}} \right)^2\right]}
    \begin{cases}
        \exp{-\frac{1}{2}\left(\frac{x}{\sigma_+} \right)^2 } & x > 0 \\
        \exp{-\frac{1}{2}\left(\frac{x}{\sigma_-} \right)^2 } & x < 0,
    \end{cases}  
\end{equation}

where $r^2 = x^2 + y^2$, $p_0$ is the pinch parameter which controls the opening angle of the lobes, $\sigma_+$ and $\sigma_-$ control the brightness of individual lobes to fit asymmetries (see \cite{cantalloubeOriginAsymmetryWind2018} for detailed discussion on WDH asymmetry), $\gamma$ controls the radial flaring of the lobes, and $N_0$ is a scale factor. To handle different orientations, we rotate the coordinate system using another configurable parameter $\theta_{PA}$ before generating the model:

\begin{equation}
    \begin{split}
        x_{PA} = \hat{y}\sin\theta_{PA} - \hat{x}\cos{\theta_{PA}}, \\ y_{PA} = \hat{y}\cos\theta_{PA} + \hat{x}\sin\theta_{PA}.
    \end{split} 
\end{equation}

Finally, we generate our models in an annulus with the outer boundary extending just beyond the AO control region which has a diameter of $24 \lambda/D$. This slight extension is necessary to combat effects stemming from the transition region between the control region and the seeing-limited portions of the image. We also excluded the innermost, high-speckle noise region (i.e., $r < 10$ pixels or $r < 120$ mas) in our model fitting and generation. To finalize the fitting region mask where the goodness-of-fit is evaluated during the MCMC, we improved our model fits by masking out the disk object. To do this, we added additional software masking in the shape of a thin, inclined ring rotated using the known position angle of the disk and the parallactic angle recorded in the headers of the image file to exclude the disk object.

\subsection{Forward-modeling through KLIP}
\label{sec:forward-modeling-klip-subsec}

\begin{table*}[h]
\centering
\footnotesize
\caption{\texttt{windsoCC}-informed $\theta_{PA}$ parameter priors.}
\vspace{6pt}
\hspace*{-2cm}
\begin{tabular}{cc|c}
\hline \hline
                  & \textbf{UT 2023 Mar 10}        & \textbf{UT 2023 Mar 13}    \\ 
                  &  $\theta_{PA}(^\circ)$ range   & $\theta_{PA}(^\circ)$ range     \\ \hline
Group 1           & (80, 110)                       & (95, 110)  \\
Group 2           & ---                            & (105, 130)  \\
Group 3           & ---                            & ---  \\
Group 4           & (100, 130)                      & (130, 150)    \\
Group 5           & (180, 210)                      & (160, 200) \\
Group 5           & ---                            & (185, 225) \\
Group 6           & (200, 230)                      & --- \\ \hline
\end{tabular}

\footnotesize
\vspace{2mm}
\raggedright \textbf{Notes.} Uniform prior probabilities.

\label{tab:priors-PA}

\end{table*}

We made use of the Python-based \textsc{DiskFM} software tool \cite{mazoyerDiskFMForwardModeling2020} which is also part of the \texttt{pyKLIP} package \citep{wangPyKLIPPSFSubtraction2015}. In this work, we have wired in our parametric model of the WDH for forward modeling through KLIP in lieue of a traditional disk model. We will refer to our application of \textsc{DiskFM} for forward modeling WDH models as \textsc{WindFM} hereafter. This analysis is needed to estimate the impact of the PSF subtraction on the WDH such that we can infer the morphology of the WDH in our individual science frames for later removal. We wrapped this forward modeling routine in an MCMC framework facilitated by the \texttt{emcee} package \cite{foreman-mackeyEmceeMCMCHammer2013}. For both datasets, we elected to fit 4 and 5 individual WDH models to our reduced \iband{} and \rband{} images, respectively. Each individual WDH model was fit using its own set of parameters (i.e., we fit $p_0$, $\sigma_+$, $\sigma_-$, $\theta_{PA}$, $N_0$ for each individual WDH model). The one exception is the flaring parameter $\gamma$ which we treated as a global parameter to combat the significant degeneracies we observed during preliminary modeling trials. Whether or not $\gamma$ should change between individual WDHs (or even be permanently fixed to some value for MagAO-X data) is an analysis we will perform in a later paper. In this way, the number of fitted parameters was 21 and 26 for our \iband{} and \rband{} images, respectively, to test if we could obtain a WDH model solution along each wind track that our \texttt{windsoCC}. To probe this hypothesis, we defined our priors for the $\theta_{PA}$ parameters centered on the wind directions detected for each dataset; see Table \ref{tab:priors-PA}. Recall, from the beginning of Section \ref{sec:forward-modeling}, for the wind vectors extracted from the UT 2023 Mar 13 dataset, we elected to fit 2 WDH models to group 4 due to its large spread of wind directions. We assumed uniform prior probability distributions for the other fitted parameters. An overview of the full forward modeling procedure is as follows:

\begin{enumerate}
    \item Generate an integer number of WDH models using the method outlined in \ref{sec:model-description-subsec}.
    \item Compute and subtract the median radial profile from this model image.
    \item Use our \textsc{WindFM} pipeline to produce a forward model (FM) through the image subtraction done using the single KLIP mode.
    \item Gauge the goodness of fit of these results using the standard $\chi^2$ metric:
    \begin{equation}
    \label{eq:chisquare}
        \chi^2 = \sum_S \frac{(\text{Data} - \text{FM})^2}{\text{Uncertainty}^2},
    \end{equation}
\end{enumerate}

\noindent where $\mathcal{S}$ is the search region where the likelihood $\mathcal{L} = e^{-\chi^2/2}$ is maximized. Step 2 is needed to ensure that the pre-processing steps that we apply to our coronagraphic image cutouts is also applied to the model image (see Section \ref{sec:data-reduction} for the details on our focal plane data reduction steps). We estimate the spatial noise map by performing KLIP but then ``back-rotating" the images by the negated parallactic angle as described in \cite{gerardPlanetDetectionFew2016} to median-out the vast majority of the disk. We then use this back-rotated KLIP image to calculate the standard deviation at every radial location using 1 pixel wide concentric measurement annuli as in \cite{lawsonSCExAOCHARISNearinfrared2020}. We note that while we were able to reliably remove the disk using this method, some residual WDH artifact remained. One additional step one could take is a spatial filtering operation to remove the remainder of the WDH before estimating the noise. One other strategy might be to apply a scale factor to the spatial noise estimate to down weight it during the model fitting, but we did not perform such experiments for this analysis.

\subsection{Removal of the wind-driven halo in individual science images}
\label{sec:removing-wdh-subsec}

After deriving the model solution using the KLIP-reduced image and \textsc{WindFM}, we used the individual WDH model components as part of a modal basis to remove the WDH in individual science frames. Namely, we built the full basis out of the WDH models that sum to make up the best-fit total model solution using \textsc{WindFM}. As an additional basis component, we constructed a median-stack of the science image library to use as a PSF estimate. It is these WDH model components and the PSF estimate that we used as part of a non-negative least-squares fitting routine to subtract the WDH artifacts in the individual science frames.

To prepare for this WDH removal procedure, we designed custom software masks to define the fitting region similarly as described above. We also added 4 small elliptical masks to remove the artificial DM speckles present in the control region (these speckles are elliptical due to the spectral width of the science camera filters; see Section \ref{sec:observations}). To finalize our masking, we excluded the disk object as explained before in Section \ref{sec:model-description-subsec} to prevent biasing the WDH model fit. Before performing the least-squares routine, we normalized each of the WDH model components using their respective L1-norm (i.e., the model component's pixel value sum equals one). We then enforced a positivity constraint with the WDH model components during fitting by using of a non-negative least squares fitting procedure. Once we derived the basis coefficients, we subtracted this scaled PSF estimate from each of the science frames.

\begin{table}[h!]
\centering
\footnotesize
\caption{WindFM Modeling Results for \hr{}}
\label{tab:windfm-results}

\begin{tabular}{lccccc}
\multirow{2}{*}{Parameter$^\text{a}$} & \multirow{2}{*}{Range$^\text{b}$} & \multicolumn{4}{c}{Best-fit values$^\text{c}$ \iband{} (UT 2023 Mar 10)} \\ \cline{3-6} 
 & & WDH 1 & WDH 2 & WDH 3 & WDH 4 \\ \hline
$p_0$                               & (0.01, 100)          & $0.507 \pm 0.013$   & $0.895 \pm 0.021$   & $2.117 \pm 0.061$   & $1.032 \pm 0.027$   \\
$\theta_{PA} (^{\circ})$            & (90, 230)$^\text{d}$ & $90.010 \pm 0.013$ & $124.996 \pm 0.005$ & $199.259 \pm 0.288$ & $224.999 \pm 0.102$ \\
$\sigma_{+}$                        & (0.1, 100)           & $24.725 \pm 0.194$  & $20.811 \pm 0.075$  & $19.673 \pm 0.198$  & $24.919 \pm 0.073$  \\
$\sigma_{-}$                        & (0.1, 100)           & $18.567 \pm 0.170$  & $20.637 \pm 0.083$  & $18.849 \pm 0.137$  & $21.698 \pm 0.067$  \\ 
$\gamma$                            & (0.7, 1.3)           & $0.738 \pm 0.003^{\text{g}}$   &  ---                & ---                 & ---                 \\ \hline
${F_{\text{WDH}} / F_{\text{Total}}}^{\text{e}}$ &                      & $0.176$             & $0.337$             & $0.196$             & $0.291$             \\ \hline
\end{tabular}

\vspace{0.5cm} 

\begin{tabular}{cccccc}
 \multirow{2}{*}{Range$^\text{b}$} & \multicolumn{5}{c}{Best-fit values \rband{}$^\text{c}$ (UT 2023 Mar 13)$^\text{f}$} \\ \cline{2-6} 
  & WDH 1 & WDH 2 & WDH 3 & WDH 4 & WDH 5 \\ \hline
 (0.01, 100)          & $0.394 \pm 0.015$   & $0.130 \pm 0.004$   & $0.055 \pm 0.002$   & $0.073 \pm 0.003$ & $0.141 \pm 0.006$ \\
 (100, 230)$^\text{d}$ & $102.755 \pm 0.197$ & $105.016 \pm 0.021$ & $142.758 \pm 0.267$ & $168.206 \pm 0.307$ & $193.653 \pm 0.437$  \\
 (0.1, 100)           & $13.515 \pm 0.130$  & $19.442 \pm 0.214$  & $12.914 \pm 0.147$  & $16.854 \pm 0.231$ & $23.968 \pm 0.342$ \\
 (0.1, 100)           & $14.343 \pm 0.101$  & $20.266 \pm 0.227$  & $13.043 \pm 0.165$  & $18.590 \pm 0.260$ & $25.537 \pm 0.371$ \\ 
 (0.7, 1.3)           & $1.029 \pm 0.005^\text{g}$   &  ---                & ---                 & ---      \\ \hline
                      & $0.294$  & $0.244$  & $0.143$  & $0.178$ & $0.142$ \\ \hline
\end{tabular}

\vspace{2mm}
\raggedright
$^{\text{a}}$ We included parameter $N_0$ in our modeling to handle the flux normalization, but it is not shown in this table. \\
$^{\text{b}}$ We assumed uniform prior distributions for all parameters. \\
$^\text{c}$ 16th, 50th, and 84th percentiles. \\
$^\text{d}$ Each prior assumed for the $\theta_{PA}$ parameter varies between each model; see Table \ref{tab:priors-PA}. \\
$^\text{e}$ We include the sum of each WDH model normalized by the total sum of all models. \\
$^\text{f}$ Row ordering is identical to the table above. \\
$^\text{g}$ Global parameter $\gamma$ affects all WDH models. \\

\end{table}

\section{\textsc{WindFM} Results: demonstration of WDH correction using data collected on-sky}
\label{sec:on-sky-demonstration}

\begin{figure}[h!]
\centering
\includegraphics[width=0.8\linewidth]{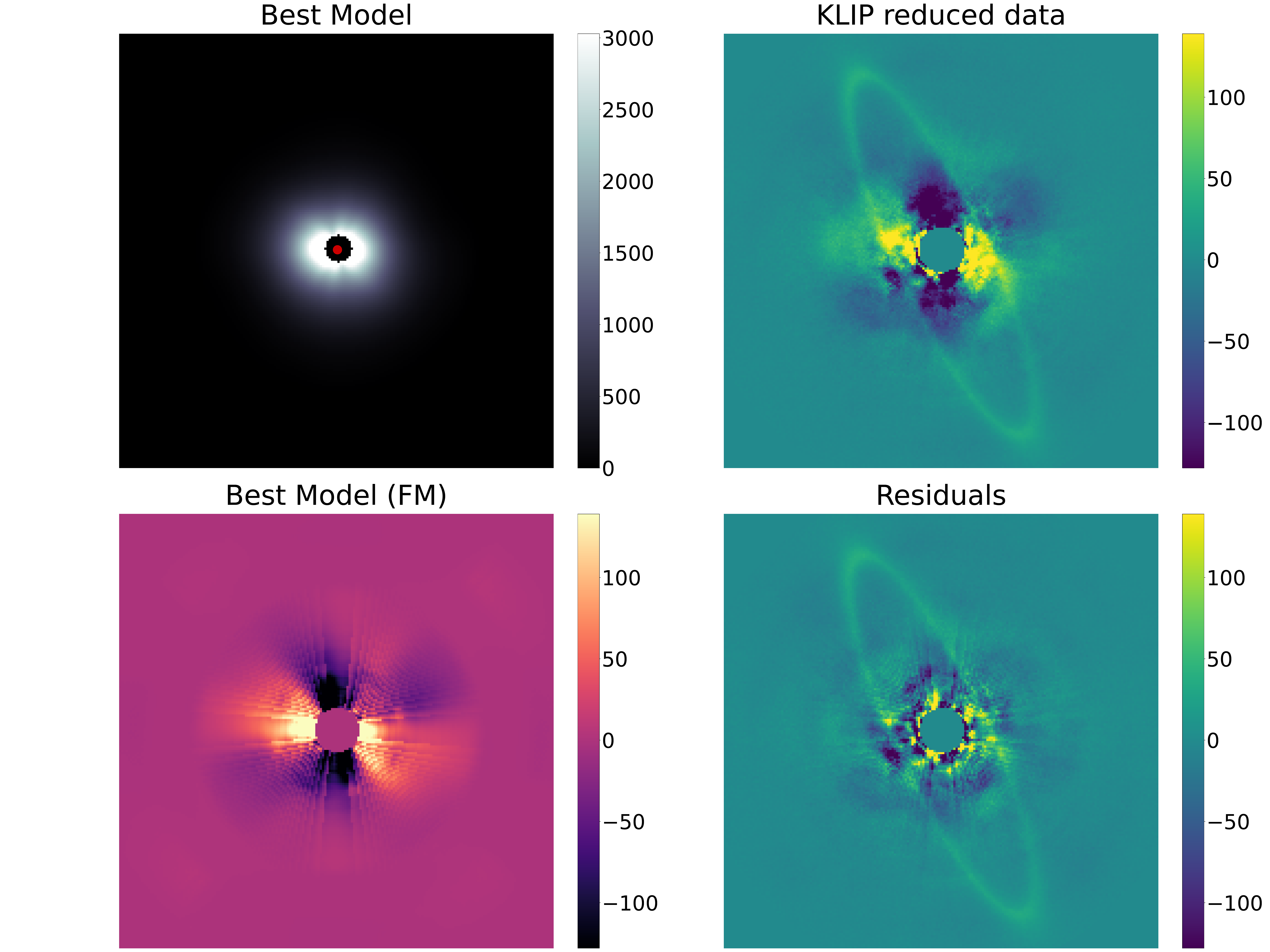}
\caption
{WDH forward modeling results using the KLIP-reduced image of \hr{} from the observations collected on UT 2023 Mar 10. \textbf{Top left:} The best-fitting model consisting of a sum of 4 individual WDH models. The location of the star behind the mask is marked by the red dot. We use the 4 WDH model components that make up this best-fit agglomerated model in our least-squares fitting routine described at the end of Section \ref{sec:removing-wdh-subsec}. \textbf{Top right:} KLIP-reduced image made using 1 KLIP mode. \textbf{Bottom left:} The best WDH model after forward-modeling through KLIP. \textbf{Bottom right:} Residuals image after subtracting the best forward model from the KLIP-reduced image.}
\label{fig:windfm-i}
\end{figure}

\begin{figure}[h!]
\centering
\includegraphics[width=0.8\linewidth]{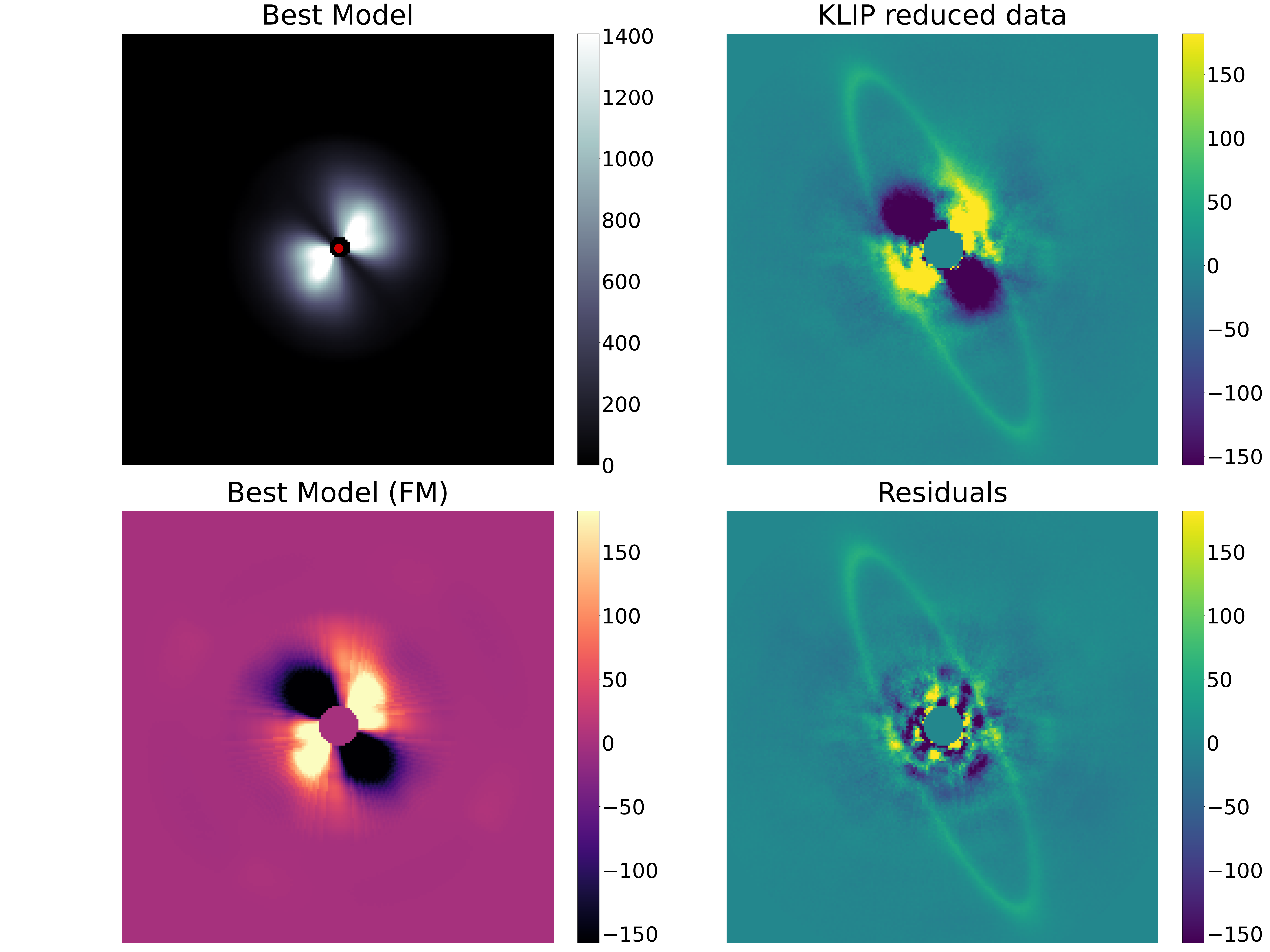}
\caption
{The same as Figure \ref{fig:windfm-i}, for the reduced image of \hr{} made from observations conducted on UT 2023 Mar 13.}
\label{fig:windfm-r}
\end{figure}

For each forward-modeling run, we iterated until the walker chains exceeded 50 autocorrelation times in length to ensure adequate sampling of the parameter space. This amounted to testing 213k and 746k models for the UT 2023 Mar 10 and UT 2023 Mar 13 datasets, respectively. The \textsc{WindFM} results pertaining to the datasets on UT 2023 Mar 10 and 13 are displayed in Figure \ref{fig:windfm-i} and Figure \ref{fig:windfm-r}, respectively. We report the best-fit values and associated uncertainties for our model parameters in Table \ref{tab:windfm-results}. We also show the range of values probed for all fitted parameters in Table \ref{tab:windfm-results} with the exception of the model flux scaling parameter $N_0$. In addition to the best-fit parameters, we supply the fractional contribution of each WDH model component by summing the pixel values in each of the individual WDH models and dividing that value by the sum of the combined WDH model (top left images in Figure \ref{fig:windfm-i} and Figure \ref{fig:windfm-r}). We report these values in the bottom rows of each sub-table in Table \ref{tab:windfm-results}. These values are informative when comparing the individual WDH model components to the wind vectors extracted from the WFS telemetry. It is worth noting that the \textsc{DiskFM} tool is known to underestimate uncertainties \cite{mazoyerDiskFMForwardModeling2020} and the exact cause of this has not yet been explored. We postulate that for this analysis, the difficulty in estimating the spatial noise map is likely affecting the posterior distributions on our fitted \textsc{WindFM} parameters.

\begin{figure}[h!]
\begin{center}
\begin{tabular}{c}
\includegraphics[width=0.95\linewidth]{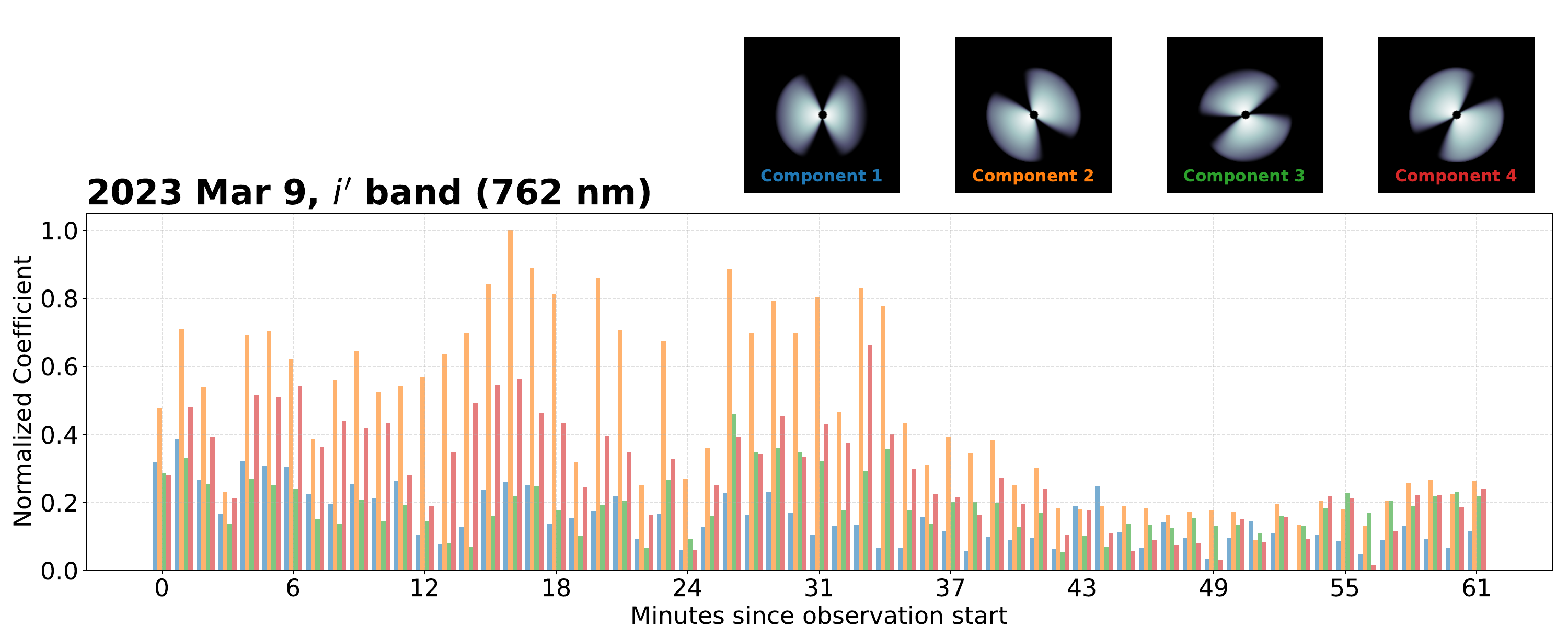} \\
\includegraphics[width=0.95\linewidth]{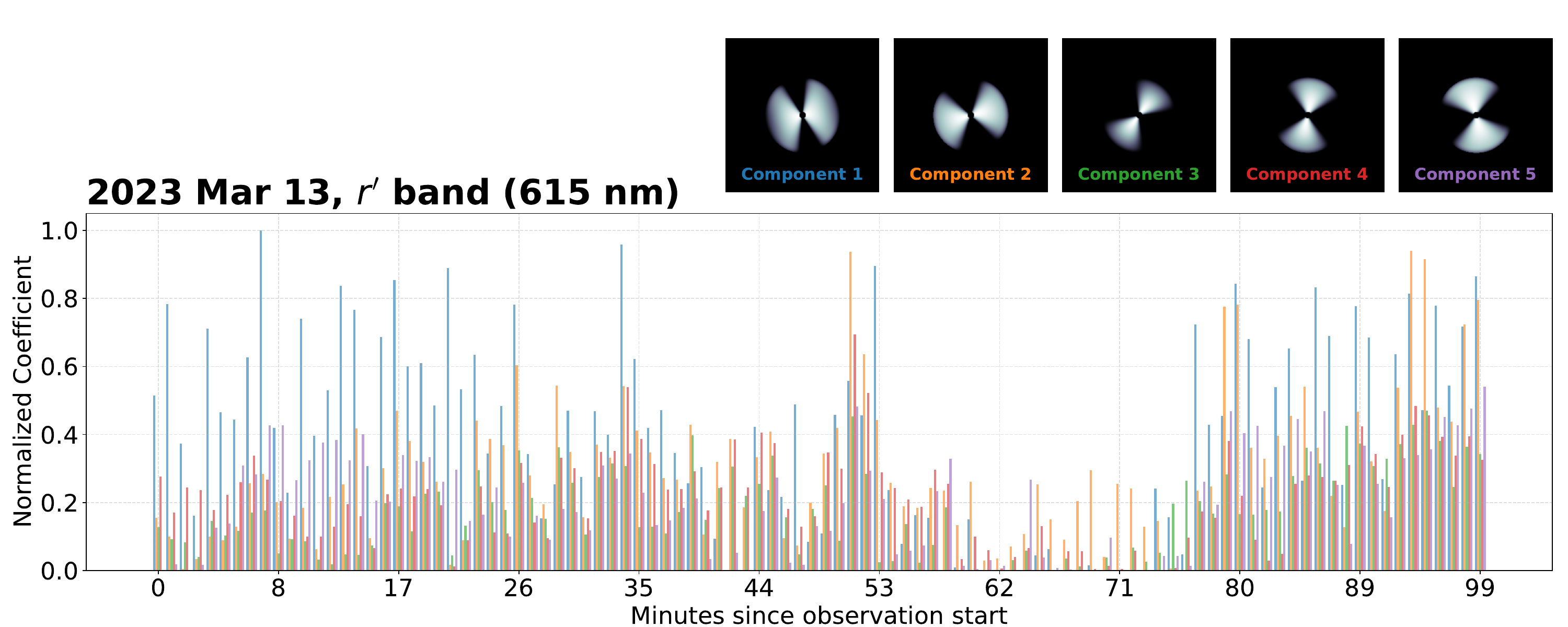}

\end{tabular}
\end{center}
\caption
{Coefficients derived from a non-negative least-squares fitting routine when including multiple WDH models in a modal basis. We show the results of this analysis using both the \iband{} and \rband{} datasets collected on UT 2023 Mar 10 and UT 2023 Mar 13, respectively. We retrieved these WDH model components from our forward modeling analysis, i.e., these components make up the best model in Figure \ref{fig:windfm-i} and Figure \ref{fig:windfm-r} when summed together. The y-axis shows the normalized coefficient value while the x-axis denotes the number of minutes elapsed since the observation started. We created this plot to show the morphology of the WDH in individual science frames and the directions of the most turbulent wind layers as a function of time.}
\label{fig:coefficients}
\end{figure}

\begin{figure}[h!]
\begin{center}
\begin{tabular}{c}
\includegraphics[width=\linewidth]{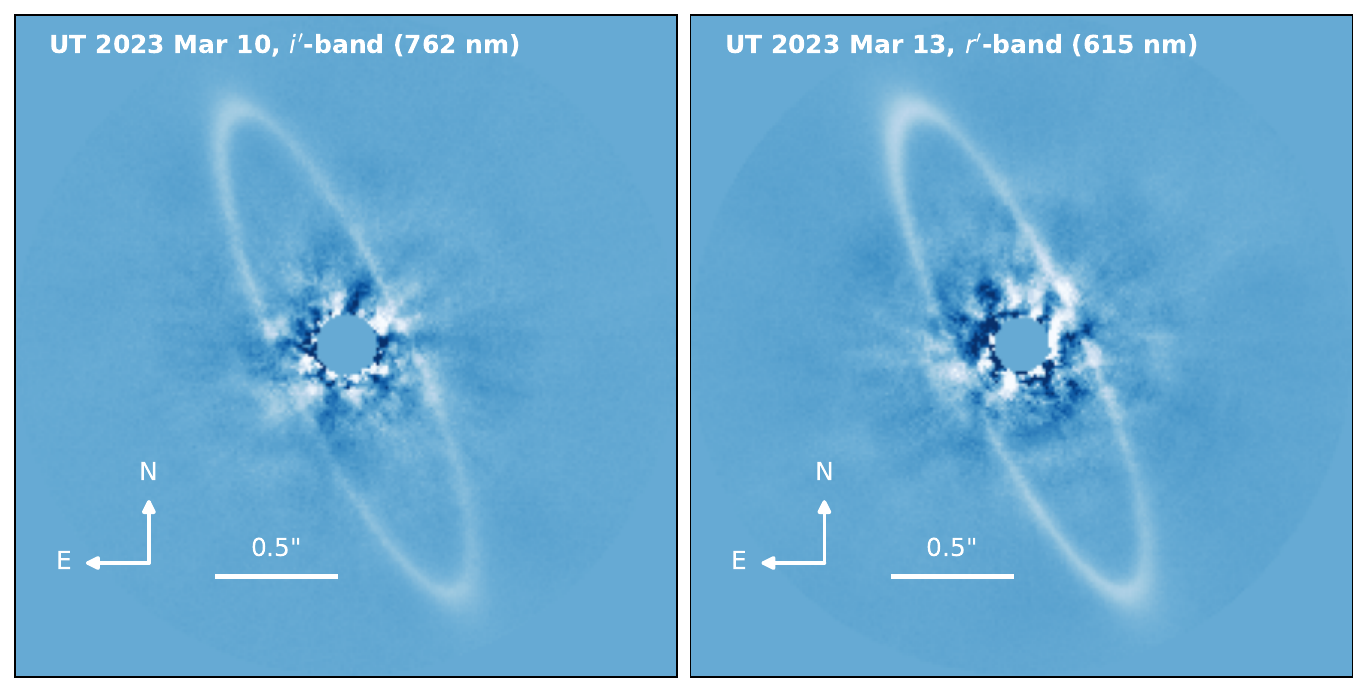}
\end{tabular}
\end{center}
\caption
{Reduced images of \hr{} after removing the WDH from individual science frames and then subtracting the PSF using a single KLIP mode. We display these images using the same scale and colormap as in Figure \ref{fig:wdh-klip}.}
\label{fig:wdhsub}
\end{figure}

After retrieving best-fit WDH model solutions using \textsc{WindFM}, we took the individual WDH model components that add up to make the total model solution (i.e., the top left panels in Figure \ref{fig:windfm-i} and Figure \ref{fig:windfm-r}) and used these components to remove the WDH artifact in individual coronagraphic science images as explained in Section \ref{sec:removing-wdh-subsec}. To retrieve the best fits, we found it necessary to perform this least-squares fit and model PSF subtraction iteratively for each dataset (or, alternatively, each temporal data cube) because the PSF estimate contains the time-averaged footprint of the WDH. Namely, with each successive iteration, we remake the median-collapsed PSF estimate in order to fully drive out the WDH artifacts. We show the results of these efforts in Figure \ref{fig:coefficients} where we used 50 iterations for our least-squares fitting routine. After attempting to remove the WDH artifacts in our individual science images using the aforementioned methods, we then made new KLIP-reduced images which we feature in Figure \ref{fig:wdhsub}. We reiterate that the bottom right panels in Figure \ref{fig:windfm-i} and Figure \ref{fig:windfm-r} are the result of subtracting off the best-fit WDH forward model from the KLIP-reduced images shown in Figure \ref{fig:wdh-klip}. Figure \ref{fig:wdhsub} shows a refreshed KLIP-reduced image after scrubbing our individual coronagraphic images of the WDH artifact. For reference, we direct readers to the second paragraph in Section \ref{sec:summary} for a high-level overview of the methods and results presented thus far.

\section{Discussion: \texttt{windsoCC}}
\label{sec:discussion-windsocc}

\subsection{Measured wind vectors using MagAO-X's pyramid WFS images}
\label{sec:discuss-wind-behavior}

In the $\sim1$ hour observing block on UT Mar 10, we observed 5 discrete clusters that were identified by the HBDSCAN algorithm (Figure \ref{fig:clusters-wind}, left side). Almost all of the wind vectors that we measured are pointing in the southerly direction, as seen in the left wind rose plot in Figure \ref{fig:directions-vs-time-wind}. Groups 5 and 6 appear to span the longest time during this observation, while groups 1, 2, and 3 were detectable in closed-loop WFS telemetry for about 10, 5, and 30 minutes, respectively. Group 1, which corresponds to the fastest layer, was detectable near the start of observations, but appears to have calmed after $\sim50$ minutes. For all groups, detections decreased significantly during the last $\sim20$ minutes of the observation. After checking the historical LCO weather records (publicly-available at: {\url{https://weather.lco.cl/stations-history}}), we believe that this is due to overdriving of the AO system after seeing that there was a spike in seeing coincident with this drop in wind detections. During the 2023A semester, MagAO-X did not have an automated way to adjust the high-order AO loop gains and instead relied on the operator's manual input. We think that in response to this sudden uptick in seeing, the AO operator increased the loop gains in an attempt to fight the worsening conditions. Our need to filter out the last quarter of the set of science images due to prominent overdriving artifacts in the AO control region is good evidence toward this conjecture (see Section \ref{sec:observations}).

The $\sim1.5$ hour dataset collected on UT Mar 13 showed a significant increase in the number of measured wind vector points and overall wind speeds (Figure \ref{fig:clusters-wind}, right side), despite similar seeing compared to the morning of UT 2023 Mar 10 (Table \ref{tab:obslog}). This is surely due to the smaller atmospheric coherence parameter $\tau_0$ (see Table \ref{tab:obslog} and, e.g., \cite{roddierAdaptiveOpticsAstronomy1999} for the definition of this parameter), as the wind at jet stream altitudes was about 6 m/s faster during these observations. As with the other dataset, the vast majority of the wind vectors that we measured using our WFS images were traveling in the southerly direction (Figure \ref{fig:directions-vs-time-wind}), consistent with historical data. We detect the wind corresponding to most groups across the entirety of the dataset with the exception of group 1 and 3 which persisted for about 1 hour and 20 minutes respectively.

A general trend seen in Table \ref{tab:wind-stats} is higher directional uncertainty associated with the lower velocity wind layers. If the trend is due to residual systematic errors, one potential cause is that the slower CC peaks are more affected by the evolving background noise. Further analysis with additional datasets from other nights would be needed to confirm or debunk this idea. Another source of systematic error may stem from the clustering analysis that we perform on the wind vectors extracted by \texttt{windsoCC} since we compute the statistics from the identified Groups in Figure \ref{fig:clusters-wind}. Indeed, as seen in Figure \ref{fig:clusters-wind}, there appear to be tight groupings of points within large clusters that could either be due to a wind layer changing direction over time, or single layers of wind being captured within these large clusters by the HBDSCAN algorithm. As explained in the next subsection, we struggled with manual HBDSCAN parameter tuning, so some clusters may be oversized. This is seen especially in the wind vector points extracted from the UT 2023 March 13 data (Figure \ref{fig:clusters-wind}, right side).

\subsection{Comparison to historical wind data}
\label{sec:discuss-compare-era5-subsec}

\begin{table}[ht]
\centering
\caption{ERA5 Atmospheric Wind Profile for LCO.}
\label{tab:era5} 
\begin{tabular}{ccccc|ccccc}
\hline\hline
\multicolumn{5}{c|}{UT 2023 Mar 10} & \multicolumn{5}{c}{UT 2023 Mar 13} \\
\hline
Pressure & Alt. & Speed & Direction & Group$^a$ &  Pressure & Alt. & Speed & Direction & Group$^a$ \\
(hPa) & (km) & (m/s) & $[^\circ]$ &  & (hPa) & (km) & (m/s) & $[^\circ]$ \\
\hline
750 & 2.5 & 0.2 & 278.5  &    & 750 & 2.5 & 1.3 & 207.0 &  \\
700 & 3.0 & 0.3 & 268.2  &    & 700 & 3.0 & 3.0 & 191.3  & 6 \\
650 & 3.6 & 0.5 & 31.0   &    & 650 & 3.6 & 4.2 & 182.9 & 6 \\
600 & 4.2 & 0.0 & 45.0   &    & 600 & 4.2 & 5.8 & 151.8 & 6 \\
550 & 4.9 & 1.4 & 217.0  & 5 & 550 & 4.9 & 7.8 & 134.9 & 5 \\
500 & 5.6 & 2.8 & 206.7  & 5  & 500 & 5.6 & 7.3 & 132.2 & 5 \\
450 & 6.3 & 2.1 & 196.9  & 5  & 450 & 6.3 & 8.6 & 136.9 & 5 \\
400 & 7.2 & 2.8 & 165.6  & 5  & 400 & 7.2 & 14.1 & 136.7  & \\
350 & 8.1 & 2.3 & 134.3  &    & 350 & 8.1 & 16.3 & 132.7 &  \\
300 & 9.2 & 1.6 & 23.9   &    & 300 & 9.2 & 22.1 & 114.8 &  \\
250 & 10.4 & 2.2 & 343.1 &    & 250 & 10.4 & 25.0 & 114.6 & 2 \\
200 & 11.8 & 9.0 & 102.9 & 2  & 200 & 11.8 & 26.6 & 104.4 & 2 \\
150 & 13.6 & 21.3 & 93.3 & 1  & 150 & 13.6 & 16.1 & 118.6 & 3 \\
100 & 16.2 & 16.0 & 94.2 & 1  & 100 & 16.2 & 10.8 & 122.4 & 3 \\
70 & 18.4 & 5.2 & 82.0   &    & 70 & 18.4 & 5.8 & 119.6 &  4 \\
50 & 20.6 & 4.2 & 285.5  &    & 50 & 20.6 & 5.8 & 260.1 & \\
\hline
\end{tabular}

\vspace{2mm}
\raggedright
$^a$ We manually assign group identification numbers to individual layers based on Figure \ref{fig:clusters-wind}.
\end{table}

To compare our measured wind vectors to historical wind data, we retrieved the Copernicus ERA5 dataset \cite{hersbachERA5GlobalReanalysis2020} which is a reanalysis of a global circulation model by the European Centre for Medium-range Weather Forecasts. We include the atmospheric wind profile above LCO returned in the ERA5 reanalysis dataset when querying for the time span UT 5:00 to 7:00 on 2023 Mar 10 and 13 in Table \ref{tab:era5}. We averaged the wind velocity and direction across this time span for brevity, as the values did not vary significantly. By inspection, because our methods do not return the altitude of the detected turbulence, we cross-matched our \texttt{windsoCC}-measured wind vectors with the ERA5 data to infer the altitude of some of wind layers negatively-affecting MagAO-X's performance as shown in columns 5 and 10. For the UT 2023 Mar 10 dataset, Groups 2 and 4 did not appear to pair with any of the ERA5 wind layers. However, these groups appear to have been present for a brief time (Figure \ref{fig:directions-vs-time-wind}) and so were likely spurious or localized gusts not captured up by any other surrounding instrumentation or wind models. This is not surprising, as the ERA5 dataset lacks the spatial resolution, temporal resolution (data are given hourly), and local meteorological input (from, e.g., ground stations, balloons, and spacecraft) required to be used as the ground truth. Our results perhaps spotlights the advantages of in-situ turbulence profiling over relying on global weather reanalysis models.

We note that we again referenced the historical LCO weather records for the ground-layer wind data, but found that these slow winds did not clearly match any one of the groups identified in our analysis. Recall that the maximum temporal delay that we used to generate our CC response maps was 2 seconds, so our response maps are not sensitive to the aforementioned slow ground-layer wind speeds during the observation. These slow wind speeds are consistent with the ERA5 predictions, see the row in Table \ref{tab:era5} for the wind properties at 2.2 km (775 hPa), which is close in altitude to the location of LCO.

With comparison to the ERA5 values, we were able to determine that Group 1 on both nights correspond to the jet stream layer. With this value, we used Eq. 5 from \cite{sarazinESODifferentialImage1990} to calculate an estimate of the effective wind velocity $V_0$. In addition, we used the median seeing values from Table \ref{tab:obslog} to compute as estimate of $\tau_0$. For both nights, we obtained $\tau_0=6.4$ ms and $\tau_0=5.0$ ms for UT 2023 Mar 10 and UT 2023 Mar 13, respectively. We append these values to Table \ref{tab:obslog} for reference. We can also use these values for $\tau_0$ to quantify the increase in servo-lag error between the two nights. If we estimate MagAO-X's high-order loop delay to be about $750~\mu$s, we can make use of $\sigma_{TD}^2=(\Delta t/\tau_0)^{5/3}$ which is a rule of thumb for determining the residual phase variance due to pure time delay (a.k.a., the servo-lag error; \cite{friedTimedelayinducedMeansquareError1990},\cite{hardyAdaptiveOpticsAstronomical1998}). Here, $\Delta t$ is the AO loop delay plus the exposure time of one WFS frame (0.5 ms here, since the loop speed was 2 kHz). Bringing this all together, we obtain $\sigma_{TD}^2 = 6.578e-2$ and $\sigma_{TD}^2 = 9.921e-2$ radians$^2$ for the nights of UT 2023 Mar 10 and Mar 13, respectively. From this quick calculation, we see that the dataset from UT 2023 Mar 13 experienced $\sim50\%$ greater phase variance due to servo-lag compared to the UT 2023 Mar 10 dataset. This is further reinforced in Figure \ref{fig:wdh-klip} where the WDH is clearly more severe in the right image.

Recall that Figure \ref{fig:wind-collpased} shows the time-collapsed CC response maps across the entire observation and so is a visualization of the stablest layers of turbulence. One interesting realization is that the jet stream layer present on the night of UT 2023 Mar 13 is weakly depicted in this visualization (right panel). This is likely due to the comparatively short window of time that this layer was exhibiting strong turbulence compared to the other layers, as shown in Figure \ref{fig:directions-vs-time-wind}, and so the tracks corresponding to the lower altitude layers dominate.

To display the content in Table \ref{tab:era5} graphically, we plot the speed and direction of the different layers over the wind vectors measured by \texttt{windsoCC} in Figure \ref{fig:clusters-wind} using light gray squares labeled with the layer altitude in kilometers. Since we assigned groups to the different layer altitudes manually, it is possible that some groups capture wind vectors from other layers of turbulence that are separated by large physical distances. For instance, in the right clustering plot in Figure \ref{fig:clusters-wind} corresponding to the observation on UT 2023 Mar 13, Group 2 appears to contain wind vector points from the layers at 13.6 km (150 hPa) and 8.1 km (350 hPa). We note that we tried reducing the distance threshold parameter used by HBDSCAN, but that caused many of the dense clusters associated with this night to be chopped up into far too many smaller clusters. Likewise, increasing the distance parameter caused the algorithm to group too many neighboring points. Since our methods dictate that we fit a WDH model for each detected cluster, reducing this distance threshold parameter would be computationally intractable due to a vast increase in the number of free parameters needing to be fit in our MCMC procedure. Despite this ambiguity in the assignments of the wind layer groups, we observed good fits with our WDH models during our \textsc{WindFM} analysis which we detail in the following section.

\section{Discussion: \textsc{WindFM}}
\label{sec:discussion_windfm}

\subsection{Best-fit wind-driven halo solutions}
\label{sec:discuss-wdh-models-subsec}

Recall that \textsc{WindFM} follows the same general workflow as the original DiskFM tool from \cite{mazoyerDiskFMForwardModeling2020}. Namely, \textsc{WindFM} assumes that the WDH is static through our observations and models the movement of it by creating $N$ copies of our WDH model and rotating each copy to assume the parallactic angle of the $N$ individual science frames. The model is then forward-modeled through KLIP. This process is performed iteratively as explained in Section \ref{sec:forward-modeling-klip-subsec} which results in the WDH model components shown in Table \ref{tab:windfm-results} (shown as a combined WDH model in the top left of Figure \ref{fig:windfm-i} and Figure \ref{fig:windfm-r}). To infer the WDH footprint in the individual science PSF image library, we fit these WDH model components in a non-negative least-squares fitting routine as detailed in Section \ref{sec:removing-wdh-subsec}. We discuss this fitting of the WDH components to individual images further in the next subsection.

Our \texttt{windsoCC} results implied that the WDH affecting our MagAO-X images is an agglomeration of multiple WDHs stemming from the turbulent layers above LCO at the time of observations. See Section \ref{sec:simulated_var_compare} for a comparison between an agglomerated WDH model and simulation results of the expected contrast performance of MagAO-X. Because we subtracted the PSF using a single KLIP mode, we assumed that the majority of the WDH remained in the reduced image since it is a dynamic object in the dataset. It is not known if every wind track identified by our \texttt{windsoCC} pipeline is responsible for a measurable contribution to the agglomerate WDH seen in the KLIP-reduced image. Questions we would like to answer in a future paper include how the stability of the layer affects its ability to be removed using our methods. For instance, for the dataset from UT 2023 Mar 10, Groups 1, 2, and 3 span roughly the same directional range but different sections in time across the dataset and distinctly-different average wind speeds. Because of these similarities in direction, we used a single WDH model component for this wind layer to avoid parameter bloat in our MCMC runs. This seemed to have produced decent results (Figure \ref{fig:windfm-i}), however the wind layers corresponding to Group 3 and 4 also did not persist for very much time.

The residuals shown in Figure \ref{fig:windfm-i} and Figure \ref{fig:windfm-r} indicate that the majority of the WDH footprint was removed from our KLIP-reduced images. The remaining noise at the minor axis of the disk or interior to that IWA is dominated by residual quasi-static speckles. However, in Figure \ref{fig:windfm-i}, there is a cross-shaped artifact visible in the best-fit FM (lower left panel) and residuals (lower right panel) that originates from the radial profile subtraction. Recall from Section \ref{sec:data-reduction} that we perform a radial profile subtraction as part of the pre-processing steps carried out on the raw data in preparation for PSF subtraction with KLIP. This artifact likely stems from the azimuthal symmetry of the best-fit WDH model (upper left panel) coupled with the way we compute the radial profile using concentric annuli of 1 pixel widths. Indeed, we use the same \textsc{WindFM} code to generate the results shown in Figure \ref{fig:windfm-r} (where the best-fitting model is not so azimuthally-symmetric), yet we do not observe this artifact affecting the residuals image to the same degree as in Figure \ref{fig:windfm-i}. In any case, the radial profile subtraction artifact does not appear to print through to the reduced images in Figure \ref{fig:wdhsub}.

As shown in Table \ref{tab:windfm-results}, some of our best-fit $\theta_{PA}$ parameters reached the boundaries of their priors. We note that we constructed these specific parameter priors to strictly span the directional range of each group identified by our clustering procedure (Figure \ref{fig:directions-vs-time-wind}), i.e., we are fitting one WDH model component for each detected layer. Indeed, we deemed this necessary as the parameter space used for our forward modeling is highly degenerate. We also find it difficult to compute an accurate spatial noise estimate for underneath the WDH in the KLIP image as there are no pixels in this region that are not occupied by disk signal or WDH signal (Figure \ref{fig:wdh-klip}). Thus, we surmise that these posterior distributions are a consequence of the degenerate parameter space and poor estimate of the underlying noise and the optimizer is attempting to fit the residual speckle noise. However, due to to our constrictive priors on $\theta_{PA}$, and given that each WDH component still assumes the characteristic butterfly shape (Figure \ref{fig:coefficients}), we believe that each model component is still fitting some portion of the total WDH signature in the \adi{} image.

\subsection{Best-fit individual model components}
\label{sec:discuss-model-components-subsec}

We can use our WDH model solutions to reveal the layers of turbulence that caused the most intense WDH in our images. The bottom rows in each sub-table in Table \ref{tab:windfm-results} report the fractional contribution of each individual WDH model component to the total agglomerated best-fit WDH model for each dataset. For the dataset from UT 2023 Mar 10, our forward modeling reveals that the strongest WDH models that assume a $\theta_{PA}$ of $\sim130^\circ$ and $\sim228^\circ$. These correspond most-closely to the Group 4 ($\bar{v}\approx4.6$ m/s) and group 6 ($\bar{v}\approx6.7$ m/s) layers (Figure \ref{fig:directions-vs-time-wind}). Interestingly, group 6 does not appear to match any of the layers in Table \ref{tab:era5} which perhaps highlights the necessity for tracking the turbulence empirically during a given observation. For the UT 2023 Mar 13 dataset, the strongest WDH model components oriented $\sim102^\circ$ and $\sim105^\circ$ likely correspond to group 1 ($\bar{v}\approx27.5$ m/s) and group 2 ($\bar{v}\approx17.6$ m/s), which most closely matches to the wind layers at jet stream altitudes.

The coefficients for each WDH model component that we derived using our least-squares fitting routine (see Section \ref{sec:removing-wdh-subsec}) in Figure \ref{fig:coefficients} may represent the prevalence of each of the WDH model components in the individual science frames on a minute-to-minute basis. One interesting observation based on these charts is that the coefficient solutions are quite variable even through short windows of time (minutes). The study by \cite{cantalloubeWinddrivenHaloHighcontrast2020} also analyzed the temporal variation of the strength of the WDH artifact in SPHERE-IFS data and reported results similar to what we show in Figure \ref{fig:coefficients}. However, we caution against over-interpreting the accuracy of the relative heights of the bars between each component because these WDH components may be fitting for other adjacent WDH structures (i.e., the components are not orthogonal and we discuss these caveats further below). Because of this, the total coefficient values for each frame in the sequence may better serve as a proxy for the total integrated strength of the turbulence above LCO through time. One could leverage these ideas to implement the \texttt{windsoCC} pipeline as a realtime tool on MagAO-X to track the wind through all the observations on a given night and automatically report this information.

For the night of UT 2023 Mar 13, the dominant WDH model component appears to the be Component 1, especially in the first quarter of the dataset (Figure \ref{fig:coefficients}). Component 1 corresponds to Group 1 in Figure \ref{fig:clusters-wind} and Figure \ref{fig:directions-vs-time-wind} which is the group inferred to be at jet stream altitudes. This is also, by-eye, the most-dominant component in the \adi{} reduced image in Figure \ref{fig:wdh-klip}. However, we also observe that Component 2 is closely aligned with Component 1 (best fit $\theta_{PA}$ values about $\sim3^\circ$ apart; see Table \ref{tab:windfm-results}). While the detected wind layers corresponding to Group 1, Group 2, and Group 3 were also traveling in similar directions (Figure \ref{fig:directions-vs-time-wind}), it is possible that the jet stream produced an exceptionally-strong and variable WDH in the science images which is being fit by both WDH Component 1 and Component 2 during the MCMC procedure. Additionally, with the exclusions explained in Section \ref{sec:data-reduction} applied, the dataset from UT 2023 Mar 10 and UT Mar 2023 13 corresponded to 61 minutes and 150 minutes of observing time, respectively. This perhaps indicates that longer exposure times allow for a stronger imprinting of the WDH on the final KLIP-reduced image, leading to a need for more aggressive PSF subtraction with KLIP.

For the night of UT 2023 Mar 10, the dominant WDH components correspond to Component 2 and Component 4 based on the values of their coefficients compared to the other two components. Component 2 is oriented along the direction of the wind layer corresponding to Group 4 in Figure \ref{fig:clusters-wind} and Figure \ref{fig:directions-vs-time-wind}. However, one caveat is that our components are not orthogonal, and so could be fitting for effects due to wind layers traveling along adjacent directions (e.g., Component 2 is fitting for Group 4, but also some of Group 1 and Group 2, etc.). One would need to conduct more work to analyze the coupling effects of a basis consisting of WDH models that overlap spatially if the goal were to fully characterize how each WDH behaves across the observations. Further, it may perhaps be useful to compare the relative values of these coefficients to the coefficients of the KLIP modes after an optimized PSF subtraction via KLIP. Deriving an optimal KLIP reduction after removing the WDH could be done via a grid search over KLIP parameters to maximize the signal-to-noise of the disk, as was done in \cite{kuenyProbingDiskIce2024}. Since the goal of our study was to demonstrate the potential of WDH removal using WFS data, we leave this sort of testing for a future paper.

\section{Comparing a \textsc{WindFM} solution to simulation}
\label{sec:simulated_var_compare}

To view the similarities and differences between our \textsc{WindFM} solutions and the expected performance at the MagAO-X focal plane given the \texttt{windsoCC}-recorded turbulence, we generated spatial variance maps assuming an ideal coronagraph \cite{cavarrocFundamentalLimitationsEarthlike2006} and pure servo-lag error. The underlying theory, physical assumptions, and recipe is outlined in \cite{malesGroundbasedAdaptiveOptics2018}. In short, we modeled the atmosphere as multiple layers of Kolmogorov phase screens translating across the telescope pupil with each layer having its own velocity and direction (i.e., frozen flow turbulence \cite{taylorSpectrumTurbulence1938}). We followed the parameters listed in Ref.~\cite{malesGroundbasedAdaptiveOptics2018}'s Table 3 to simulate the hardware constraints associated with the Magellan-Clay telescope and MagAO-X. We computed the servo-lag error using the AO loop parameters, overall Fried parameter $r_0$, the velocity of each wind layer, and their relative strengths (i.e., some fractional number for each layer such that all layer strengths add to 1).  We then compute the power spectral density (PSD) of the residual phase across the pupil due to this servo-lag. Finally, we generate the aforementioned variance map via the integral of this temporal PSD after applying optimal modal integrator control.

\begin{table}[]
\caption{Wind layer information used for simulating our variance map for UT 2023 Mar 13 07:15.}
\centering
\begin{tabular}{lccc}
\hline
        & Relative Strength & Direction  & Velocity \\
        & ($\%$)            & ($^\circ$) & (m/s)    \\ \hline
Layer 1 & 10.4              & 116.2      & 27.8     \\
Layer 2 & 37.8              & 133.1      & 12.1     \\
Layer 3 & 19.9              & 141.9      & 8.1      \\
Layer 4 & 23.3              & 225.6      & 6.4      \\
Layer 5 & 8.6               & 201.1      & 4.5      \\ \hline
\end{tabular}
\label{tab:wind_for_variance_sim}
\end{table}

For a specific variance map estimation at a certain time, we referenced the \texttt{windsoCC}-extracted wind vectors corresponding to UT 2023 Mar 13 07:15; this time corresponds to the end of the $r'$-band dataset (Table \ref{tab:obslog}). We felt that this was an interesting moment to analyze given that the turbulence was strong and seemingly increasing (Figure \ref{fig:coefficients}, bottom plot). The strong layers of turbulence detected by \texttt{windsoCC} along with their relative strength, speed, and direction at this time are recorded in Table \ref{tab:wind_for_variance_sim}. We determined the relative strengths in these layers by performing aperture photometry on the CC peaks using circular apertures with a radius of 3 pixels. Namely, we assume that the CC response should be directly proportional to the refractive index structure parameter $C_n^2$. We summed each peak separately in different frames within the CC response cube such that the peaks were at a distance of 18 pixels from the image center (equivalently, the wind layer travels 2 meters in the pupil before we analyze its CC response). The reason for this is because some CC peaks travel in similar directions at similar speeds so some time must elapse to allow them to separate for accurate aperture sums (see Figure \ref{fig:tracking-wind-demo}).

\begin{figure}[h!]
\centering
\includegraphics[width=0.9\linewidth]{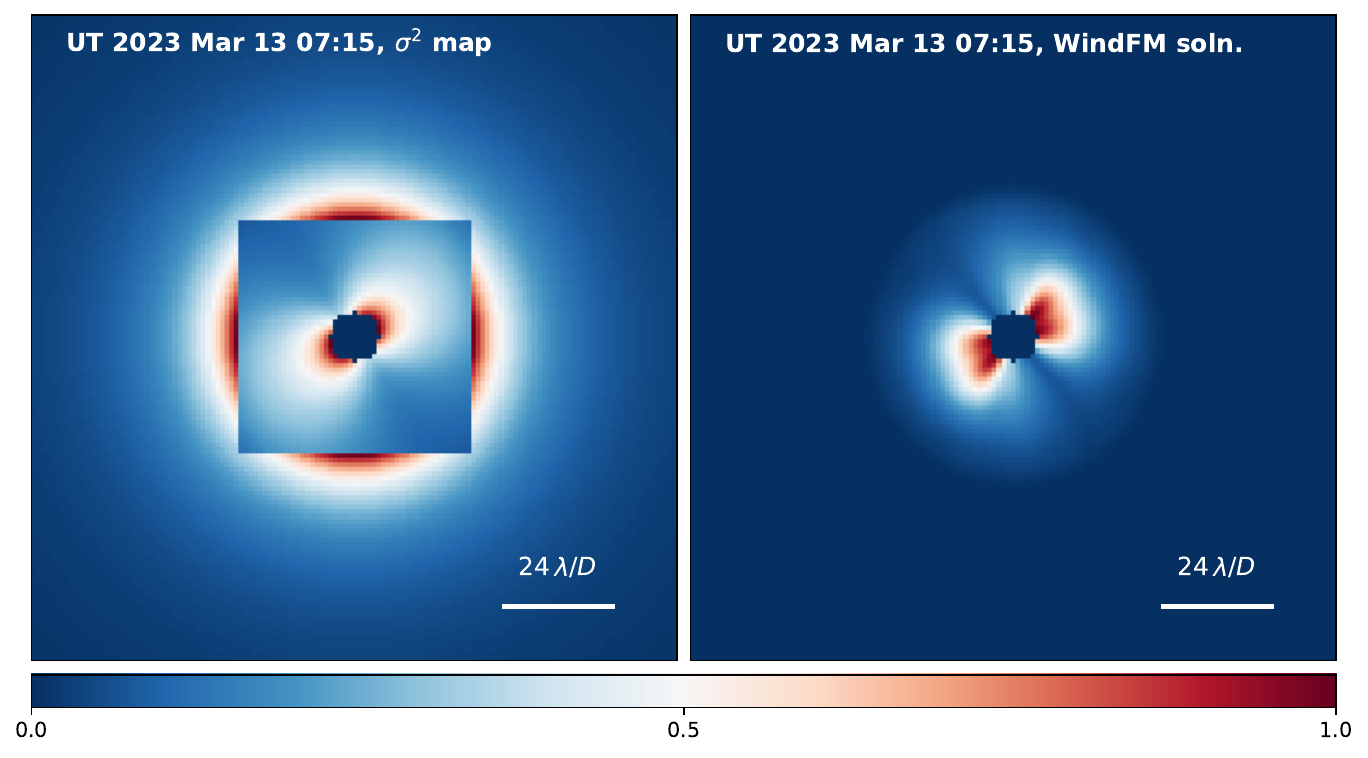}
\caption
{Comparison of the simulated variance map assuming pure servo-lag (left) with our \textsc{WindFM} solution (right). The \textsc{WindFM} solution is the WDH model components that make up the top left panel in Figure \ref{fig:windfm-r} fitted to match the WDH footprint in an individual science frame recorded at UT 2023 Mar 13 07:15. We fit the WDH model components to the science image using the least-squares fitting methods explained in Section \ref{sec:removing-wdh-subsec}. We used the turbulence parameters shown in Table \ref{tab:wind_for_variance_sim} to compute the variance map. MagAO-X's control region has a radius of $24~\lambda/D$ which we illustrate with the white bar. Each image is normalized such that the maximum value within the control region is one. We mask the innermost $5~\lambda/D$ which is consistent with the IWA  used during this analysis.}
\label{fig:variance_sim}
\end{figure}

To obtain a representative \textsc{WindFM} solution, we used the coefficient values corresponding to the last frame in the data sequence (UT 07:15 on 2023 Mar 13; see Figure \ref{fig:coefficients}) to combine the WDH components to best approximate the WDH footprint at this time. The reason that we did not use the best-fit \textsc{WindFM} model shown in Figure \ref{fig:windfm-r} (top left) to compare to a variance map made from the wind attributes averaged over the entire dataset (i.e., using the information contained in Table \ref{tab:wind-stats} to calculate a variance map) is because of the difficulty currently involved. Namely, automatically tracking the relative strengths in the different turbulent layers across the entire dataset requires complex operations and validation testing that we are choosing to leave for future work. In any case, we show the comparison between the \textsc{WindFM} solution and our variance map for UT 2023 Mar 13 07:15 in Figure \ref{fig:variance_sim}. Since this is for a high-level comparison only, we normalized both images such that the max value is unity within the AO control region and masked out the innermost $5~\lambda/D$. For the variance map (left), the control region is the square area in the center.

Through visual comparison, and considering the morphology within the AO control region, we note similar features between these two WDH signatures. The most striking difference that we observe is the seemingly multi-lobed aspect of the WDH in the \textsc{WindFM} solution. It is not known how accurately this captures the true, real-world raw contrast performance of MagAO-X considering servo-lag error or if this is a result of residual systematics. As such, more datasets need to be subject to this sort of analysis before we can make any strong conclusions on this phenomenon. Recall, that we optimize the combination of the WDH components in a least-squares fashion to the individual coronagraphic PSF images. It is possible that this fitting routine is incurring a bias from close-in quasistatic speckles. One other cause might be stellar leakage due to jitter also at the edge of the IWA. 

\section{Summary \& future work}
\label{sec:summary}

We observed \hr{} using MagAO-X at \gband{} (527 nm), \rband{} (615 nm), \iband{} (762 nm), and \zband{} (908 nm) on UT 2023 Mar 10 and 13. We detected the disk at high signal-to-noise ratios but our observations (and thus our KLIP-reduced images) were heavily affected by WDH artifacts. The goal of this study was to remove these WDH artifacts from our science data using the WFS telemetry and WDH modeling. We continue below with a brief high-level summary of the methods and results we presented in this paper.

Since MagAO-X also saves its pyramid WFS images in addition to the science data, we built a WFS image processing pipeline based on the cross-correlation (\texttt{windsoCC}) to measure the wind attributes over LCO during the observation. Namely, \texttt{windsoCC} cross-correlates the pupil images in the WFS data with pupil images at a later time to record the speed and direction of the turbulent wind layers. For our hours-long observing sequences, we found that performing a simple clustering analysis on the extracted wind vectors revealed many distinct layers of turbulence affecting our data. We then used this information to decide on the number of WDH models and the parameter priors to use for our forward modeling (for example, for the night of UT 2023 Mar 13, we used 5 WDH models during our forward modeling with their $\theta_{PA}$ priors all roughly constrained to the southerly direction based on the \texttt{windsoCC} results). We used the DiskFM tool \cite{mazoyerDiskFMForwardModeling2020} to facilitate KLIP forward modeling; we refer to our implementation as \texttt{WindFM} since we are using WDH models in lieu of traditional disk models. Since the WDH artifacts were so pronounced in our KLIP-reduced images of \hr{}, we used these reduced images to infer the WDH artifacts plaguing the science focal plane data. Next, we forward modeled our parametric WDH models using \texttt{WindFM} and demonstrated successful retrieval of extended structures composed of an agglomeration of WDHs. Finally, to scrub our individual science images of these WDHs, we took the WDH model components that make up our \texttt{WindFM} solutions and used them to assemble a modal basis so that we could subtract off the WDH footprint in each individual science image through a least-squares fitting routine. We showed that our methods facilitated images of the entire disk without the need for aggressive PSF subtraction. Though we only include results for 1 out of the 2 bandpasses for each night for brevity, the WDH solution for the second bandpass for each night is identical after applying a simple $\lambda/D$ spatial scaling adjustment.

During our analysis we demonstrated:

\begin{itemize}
    \item Retrieval of wind vectors using archival closed-loop pyramid WFS telemetry in an automated fashion. (\texttt{windsoCC})
    \item KLIP forward modeling of multiple WDHs informed by the wind vectors returned by our WFS data processing pipeline. (\texttt{WindFM})
    \item Successful removal of the WDH from MagAO-X science images using our best-fitting forward model solution scaled for each image using a non-negative least squares fit.
\end{itemize}

As we have seen, the number of turbulent wind layers affecting AO performance throughout an observation can be large. Going forward, we hypothesize that forward modeling a WDH for all wind layers detected using closed-loop WFS images will lead to a better understanding of how the current atmospheric profile will affect observations. However, for large numbers of layers, the number of free parameters to fit becomes cumbersome. One way to approach this problem is with automatic differentiation (``autodiff") which can trivialize the number of free parameters needed for fitting dozens of parametric WDH models with \textsc{WindFM}. One planned upgrade for this project is computing the forward modeling procedure within the framework of \textsc{Jax} \cite{bradburyJAXComposableTransformations2018}, which is a Python machine learning library that facilitates autodiff and GPU-accelerated array computations. \textsc{Jax} has been implemented in a KLIP forward modeling paradigm previously with encouraging results \cite{kuenyFreeformForwardmodelingPipeline2026}.

The disk around \hr{} possesses the ideal morphology for our WDH subtraction techniques described in this paper. However, there are other disk morphologies that will undoubtedly make disentangling the WDH with the astrophysical signal more difficult. Given this scenario, one could perhaps make use of the star-hopping method \cite{wahhajSearchFifthPlanet2021} (which is an application of the reference differential imaging technique; \cite{ruaneReferenceStarDifferential2019}) and use the library of reference PSFs to derive the basis of WDH models to subtract it from the target images.

Though the ideas in the preceding paragraph have potential, the ultimate goal for this project is to remove the WDH artifacts in the science focal plane images using solely the WFS telemetry. This would involve simulating the speckle pattern at the science focal plane using the information encoded in the CC-response maps measured by \texttt{windsoCC}. In Section \ref{sec:removing-wdh-subsec}, we explained how we use prior knowledge of the disk morphology to create a custom software mask to remove the disk signal from the model fitting region in the KLIP-reduced image. It is worth noting that WDH reconstruction driven purely by the WFS telemetry would also eliminate the need for this prior knowledge.

The \texttt{windsoCC} pipeline is an effective tool for informing the presence of the WDH in our coronagraphic images. Though we created this pipeline with the intention to use it during post-processing, a realtime \texttt{windsoCC} implementation on MagAO-X during observations would be very useful. In particular, knowing the state of the atmospheric turbulence could help inform target selection and facilitate more efficient use of the Magellan-Clay telescope (e.g., perhaps the observer will defer observations of a very diffuse and faint disk if the WDH is forecasted to be exceptionally strong by \texttt{windsoCC}). Moreover, the CC peaks that are teased out by the pipeline using the closed-loop residuals have the potential to promote realtime calculations of the atmospheric coherence time parameter $\tau_0$ and the profile of the refractive index structure constant $C_n^2$ (see, e.g., \cite{guesalagaUsingC2Wind2014}, \cite{sivoAutomaticWindSpeed2018} for more on these ideas). Knowledge of these parameters will be invaluable for monitoring the nightly turbulence profile above LCO and also for further automation the high-order control loop gains on MagAO-X.

\subsection*{Disclosures}

A portion of the codebase used to facilitate the analyses presented in this maniscript were developed with the assistance of artificial intelligence (AI). This research has made use of the Cursor integrated development environment (IDE) which specializes in facilitating coding assistance with large language models (LLMs) through natural language prompts. We utilized coding ``agents" to accelerate development by delegating boilerplate code generation, documentation, and reformatting. Our contributions to the codebase include the design of the algorithms, logic, and use philosophies. All AI-assisted coding operations were performed using a human-in-the-loop methodology. We have personally verified that all of the code in the latest version generates output that is correct to the best of the authors' knowledge and abilities.

\subsection*{Code and Data Availability}

The code that we used in our analysis are available upon request from the corresponding author.

\subsection*{Acknowledgments}

J.K.K., J.R.M., Y.D.L., and A.J.W. acknowledge support from the NSF, grant no. AST-2307613. J.D.L. was supported by the Flatiron Software Research Fellowship at the Flatiron Institute, a division of the Simons Foundation.

This research contains modified Copernicus Climate Change Service information [2026]. Neither the European Commission nor ECMWF is responsible for any use that may be made of the Copernicus information or data it contains.

We are very grateful for support from the NSF MRI Award No. 1625441. The Phase II upgrade program is made possible by the generous support of the Heising-Simons Foundation. MagAO-X uses the CACAO software package, which is supported by NSF Award No. 2410616.

J.K.K. and J.R.M. acknowledge support from the NSF ATI Grant No. 2308351 for PSF reconstruction.

This material is based upon High Performance Computing (HPC) resources supported by the University of Arizona TRIF, UITS, and Research, Innovation, and Impact (RII) and maintained by the UArizona Research Technologies department.

\ This research has made use of SAO Image DS9, developed by Smithsonian Astrophysical Observatory \cite{joyeNewFeaturesSAOImage2003}.

\ This work made use of the \texttt{astropy} \citep{astropycollaborationAstropyCommunityPython2013}, \texttt{SciPy} \citep{scipy1.0contributorsSciPy10Fundamental2020}, and \texttt{scikit-learn} \cite{scikit-learn} software libraries.


\bibliography{windsocc}   
\bibliographystyle{spiejour}   

\listoffigures
\listoftables

\end{spacing}
\end{document}